\documentclass[preprintnumbers, floatfix,letterpaper,aps,prd,epsfig,nofootinbib,
longbibliography,
twocolumn
]{revtex4-1}
\usepackage{bm,graphicx,dcolumn,epstopdf,epsf, latexsym,mathbbol, amssymb,amsmath,color,slashed, mathrsfs,mathcomp,simplewick}
\usepackage[center]{subfigure}
\usepackage{multirow}
\usepackage{makecell}
\usepackage[colorlinks,linkcolor=blue,citecolor=blue,urlcolor=blue]{hyperref}

\begin{document}

\allowdisplaybreaks
 \newcommand{\bq}{\begin{equation}}
 \newcommand{\eq}{\end{equation}}
 \newcommand{\bqn}{\begin{eqnarray}}
 \newcommand{\eqn}{\end{eqnarray}}
 \newcommand{\nb}{\nonumber}
 \newcommand{\lb}{\label}
 \newcommand{\f}{\frac}
 \newcommand{\p}{\partial}
\newcommand{\PRL}{Phys. Rev. Lett.}
\newcommand{\PLB}{Phys. Lett. B}
\newcommand{\PRD}{Phys. Rev. D}
\newcommand{\CQG}{Class. Quantum Grav.}
\newcommand{\JCAP}{J. Cosmol. Astropart. Phys.}
\newcommand{\JHEP}{J. High. Energy. Phys.}
\newcommand{\red}{\textcolor{black}}
%
\title{Slowly rotating black hole in chiral scalar-tensor theory}

\author{Ze-Kai Yu${}^{a, b}$}
\email{211122090023@zjut.edu.cn}

\author{Lei Liu${}^{a, b}$}
\email{liuleizjut@zjut.edu.cn}

\author{Tao Zhu${}^{a, b}$}
\email{Corresponding author: zhut05@zjut.edu.cn}


\affiliation{${}^{a}$Institute for Theoretical Physics \& Cosmology, Zhejiang University of Technology, Hangzhou, 310023, China\\
${}^{b}$ United Center for Gravitational Wave Physics (UCGWP),  Zhejiang University of Technology, Hangzhou, 310023, China}

\date{\today}

\begin{abstract}

The chiral scalar-tensor theory is an extension of the Chern-Simons modified gravity by introducing couplings between the first and second derivatives of the scalar field and parity-violating spacetime curvatures. A key feature of this theory is its explicit breaking of parity symmetry in the gravitational sector, which is expected to affect the spatial-time component of axisymmetric spacetime. In this paper, we investigate the effects of the chiral scalar-tensor theory on slowly rotating black holes by building on known solutions in the dynamical Chern-Simons modified gravity. Using perturbative methods with small coupling and slow rotation approximations, we find that the contributions of the chiral scalar-tensor theory appear at quadratic order in the spin and cubic order in the coupling constants. Furthermore, we explore the properties of this solution in the weak field and check its ergosphere and horizon. In the weak limit, we find that the effects of parity violation are suppressed in the weak field but could become significant in the strong field regime. These results provide insights into the behavior of parity-violating gravity in the presence of rotation and may be used for further investigations into its observational signatures.

\end{abstract}


\maketitle
\section{Introduction}
\renewcommand{\theequation}{1.\arabic{equation}} \setcounter{equation}{0}

Einstein's theory of general relativity (GR) has stood as the most successful framework for understanding gravitational interactions for over a century. Its theoretical predictions have been repeatedly validated through a diverse array of experimental and observational tests. In the weak-field regime, where gravitational effects are relatively small, GR's theoretical predictions have been confirmed with exacting precision, encompassing phenomena such as the precession of Mercury's perihelion \cite{Park:2017zgd}, the bending of starlight near the Sun \cite{Fomalont:2009zg, Shapiro:2004zz}, and the gravitational redshift of light \cite{Will:2014kxa}. While in the strong-field domain, GR has continued to withstand intensive scrutiny, with observations of binary pulsar systems \cite{Stairs:2003eg, Wex:2014nva, Kramer:2021jcw}, the imaging of the supermassive black holes in the M87* galaxy \cite{EventHorizonTelescope:2019dse} and Sgr A* in the galactic center \cite{EventHorizonTelescope:2022wkp}, and the direct detection of gravitational waves all lending robust support to its theoretical foundations \cite{LIGOScientific:2016aoc, KAGRA:2021vkt, LIGOScientific:2016vbw}.

Black hole is one of the most indispensable predictions of GR. Its existence has been confirmed in a lot of remarkable observations \cite{EventHorizonTelescope:2019dse, EventHorizonTelescope:2022wkp, LIGOScientific:2016aoc, KAGRA:2021vkt, LIGOScientific:2016vbw, Ghez:2008ms}, which have further cemented GR's status as the preeminent theory of gravity. In recent years, the availability of detection data on black holes has increased significantly \cite{EventHorizonTelescope:2019dse, EventHorizonTelescope:2022wkp, LIGOScientific:2016aoc, KAGRA:2021vkt, LIGOScientific:2016vbw}, making the study of black holes a topic of growing interest in the scientific community. In GR, an astrophysical black hole is expected to be described by the Kerr spacetime, characterized solely by its mass and angular momentum. This remarkable prediction has found remarkable corroboration in a wealth of recent observations, including the detection of gravitational waves \cite{LIGOScientific:2016aoc, KAGRA:2021vkt, LIGOScientific:2016vbw}, black hole images \cite{EventHorizonTelescope:2019dse, EventHorizonTelescope:2022wkp}, and the stars orbiting the supermassive black hole in our Galactic Center \cite{Ghez:2008ms}. Future precise observations, such as gravitational wave detection and black hole images, can provide more accurate and details information about the nature of the black hole spacetime in the regime of strong gravity. Importantly, these precise observations can also provide a significant way to probe or constrain possible deviations beyond Kerr spacetime.

On the other hand, symmetry is a cornerstone of fundamental theories in modern physics and plays a crucial role in the development of the 20th-century physics. Consequently, testing the symmetry of the fundamental theories through precise experiments and observations is of significant importance. It was well established early in the 1950s that one of the discrete symmetries, parity, is broken in the weak interaction \cite{Lee:1956qn, Wu:1957my}, revealing that our nature is parity violating. This naturally raises an intriguing question: could parity symmetry also be broken in the gravitational sector? Recently, a lot of theoretical models of gravity with parity violation have been studied in the literature. These include models such as Chern-Simons gravity \cite{Jackiw:2003pm, Alexander:2009tp}, chiral scalar-tensor theories \cite{Crisostomi:2017ugk}, parity-violating theory in the symmetric teleparallel gravity \cite{Conroy:2019ibo, Li:2022vtn, Li:2021mdp}, Ho\v{r}ava-Lifshitz gravity \cite{Horava:2009uw, Wang:2012fi}, Nieh-Yan modified teleparallel gravity \cite{Li:2021wij}, spatial covariant gravities with parity violations \cite{Yu:2024drx, Hu:2024hzo, Gao:2019liu}. 
Moreover, in some cases, parity violation in the gravitational sector appears to emerge as an almost inevitable feature of fundamental theories, such as string theory and loop quantum gravity \cite{Alexander:2009tp}. This growing theoretical interest has motives extensive efforts to test parity symmetry of gravity, both in the weak-field and strong-field regimes, through experimental and observational investigations, for examples, see \cite{Wang:2020cub, Zhao:2019szi, Zhao:2019xmm, Qiao:2019wsh, Yunes:2008ua, Ali-Haimoud:2011wpu, Seto:2006hf, Alexander:2017jmt, Yagi:2017zhb, Yunes:2010yf, smith2008, Wang:2025fhw, Li:2024fxy, Lin:2023npr, Zhu:2023rrx, Qiao:2022mln, Li:2022grj, Zhu:2022uoq, Zhu:2022dfq, Gong:2021jgg, Wu:2021ndf} and references therein.

Parity violation in the gravitational sector introduces distinctive corrections to the spacetime metric, particularly affecting the spatial-time components of spacetime around gravitational sources. In the weak-field regime, this manifests as modifications to the gravitomagnetic sector of spacetime predicted by GR. For instance, in (non-)dynamical Chern-Simons gravity \cite{Alexander:2007zg, Alexander:2007vt, Smith:2007jm}, ghost-free parity-violating theory \cite{Qiao:2023hlr, Qiao:2021fwi}, and Nieh-Yan modified teleparallel gravity \cite{Rao:2021azn}, parity violation gives rise to additional terms in the vectorial sector of the metric within the weak-field, post-Newtonian approximation. These corrections are proportional to the coupling constant of the corresponding theories and the curl of the post-Newtonian vector potentials \cite{Alexander:2007zg, Alexander:2007vt}. In dynamical Chern-Simons gravity, parity-violating effects typically emerge in the spatial-time components of the metric around slowly rotating black holes \cite{Yunes:2009hc, Cardenas-Avendano:2018ocb, Yagi:2012ya, Ali-Haimoud:2011zme, Konno:2009kg}. Although the specific features of parity violation may vary across different theories, a common outcome is the introduction of novel contributions to frame-dragging effects caused by rotating bodies. Consequently, measurements of frame-dragging effects in solar system experiments, such as those conducted with the LAGEOS and LARES satellites \cite{Ciufolini:2019ezb}, the Gravity Probe B mission \cite{Everitt:2011hp}, serve as valuable tools for constraining parity-violating effects in various gravitational models \cite{Qiao:2021fwi, Qiao:2023hlr}. The frame-dragging effects associated with the orbital precessions of the double/binary pulsars \cite{Kramer:2021jcw, VenkatramanKrishnan:2020pbi} offer another promising avenue for investigation \cite{Yagi:2013mbt, Yunes:2008ua, Ali-Haimoud:2011wpu}. In addition, the Lense-Thirring effect induced by the spin of black holes have been explored in several astrophysical systems \cite{Miller-Jones:2019zla, Cui:2023uyb, Hannam:2021pit, Islam:2023zzj, Li:2023lqz, Li:2022grj}. With advancements in observational techniques, current and future studies of the binary pulsars and black holes may provide critical insights into the nature of parity symmetry in gravity.

To exploring the parity-violating effects in the astrophyscial system, it is essential to first derive the parity-violating corrections to the Kerr spacetime. Obtaining rotating black hole solutions of parity-violating theories, espscially those with higher derivatives, is a very complicated task, and for that reason only approximate  solutions or numerical ones are known. This has been largely explored in a typical parity-violating gravity, the dynamical Chern-Simons gravity, in which the slowly rotating black holes have been obtained both perturbatively \cite{Konno:2007ze, Ali-Haimoud:2011zme, Yagi:2012ya, Cardenas-Avendano:2018ocb, Cano:2019ore} and numerically \cite{Delsate:2018ome}. In this paper, based on the slowly rotating black hole solutions in the Chern-Simons gravity, we derive the slowly rotating black hole solutions in the  chiral scalar-tensor theory. We observe that the contributions of the chiral scalar-tensor theory to the solution are in the quadratic order in the spin and cubic order in the coupling constants. The properties of this solution are also explored.

This paper is organized as follows. In the next section, we provide a brief overview of the chiral scalar-tensor theory and its basic field equations. Section III begins by describing the approximations used to find the black hole solutions in this theory and then continues to describe the new solution found in this paper.  In Section IV, we investigate the basic properties of this new black hole solution, including its horizon, innermost stable circular orbit, ergosphere, etc. Finally, the conclusions and summary are provided in Section V.

Throughout this paper, we adopt the metric convention $(-,+,+,+)$, with Greek indices $(\mu, \nu, \dots)$ running over $0,1,2,3$ and Latin indices $(i, j, k)$ running over $1,2,3$. Natural units are used, setting $\hbar = c = 1$.

\section{Chiral scalar-tensor theory}
\renewcommand{\theequation}{2.\arabic{equation}} \setcounter{equation}{0}

In this section, we present a brief introduction of the chiral scalar-tensor theory proposed in \cite{Crisostomi:2017ugk}. The action of chiral scalar-tensor gravity has the form
\bqn\lb{action}
S &=& \int d^4 x \sqrt{-g}\left(\frac{R}{16\pi G}+\mathcal{L}_{\rm PV}+\mathcal{L}_{\phi} + \mathcal{L}_{\rm m}\right),
\eqn
where $G$ is the gravitational constant, $R$ is the Ricci scalar, $\mathcal{L}_{\rm PV}$ is a parity-violating Lagrangian, $\mathcal{L}_\phi$ is the Lagrangian for scalar field, which is coupled non-minimally to gravity, and ${\cal L}_{\rm m}$ denotes the Lagrangian of the matter field. Note that we also use $\kappa_g \equiv 1/(16\pi G)$ in this paper. As a simplest example, we consider the action of the scalar field as
\bqn
\mathcal{L}_\phi =-\frac{1}{2} g^{\mu \nu} \partial_\mu \phi \partial_\nu \phi -V(\phi).
\eqn
Here $V(\phi)$ denotes the potential of the scalar field. The parity-violating Lagrangian of the theory can be written in the form
\bqn\lb{PV_L}
\mathcal{L}_{\rm PV} = \mathcal{L}_{\rm CS} + \mathcal{L}_{\rm PV1} + \mathcal{L}_{\rm PV2},
\eqn
where the Chern-Simons term $\mathcal{L}_{\rm CS}$ is given by
\bqn
\mathcal{L}_{\rm CS} = \frac{1}{4}\vartheta(\phi) \;^*R R,
\eqn
with
\bqn
\;^*R R=\frac{1}{2} \varepsilon^{\mu\nu\rho\sigma} R_{\rho\sigma \alpha\beta} R^{\alpha \beta}_{\;\;\;\; \mu\nu}
\eqn
being the Pontryagin density and $\varepsilon^{\rho \sigma \alpha \beta}$ the Levi-Civit\'{a} tensor defined in terms of the the antisymmetric symbol $\epsilon^{\rho \sigma \alpha \beta}$ as $\varepsilon^{\rho \sigma \alpha \beta}=\epsilon^{\rho \sigma \alpha \beta}/\sqrt{-g}$ and the CS coupling coefficient $\vartheta(\phi)$ being an arbitrary function of $\phi$. 

$\mathcal{L}_{\rm PV1}$ contains the first derivative of the scalar field and is given by
\bqn
\mathcal{L}_{\rm PV1} &=& \sum_{\mathcal{A}=1}^4  a_{\mathcal{A}}(\phi, \phi^\mu \phi_\mu) L_{\mathcal{A}},\\
L_1 &=& \varepsilon^{\mu\nu\alpha \beta} R_{\alpha \beta \rho \sigma} R_{\mu \nu\; \lambda}^{\; \; \;\rho} \phi^\sigma \phi^\lambda,\nonumber\\
L_2 &=&  \varepsilon^{\mu\nu\alpha \beta} R_{\alpha \beta \rho \sigma} R_{\mu \lambda }^{\; \; \;\rho \sigma} \phi_\nu \phi^\lambda,\nonumber\\
L_3 &=& \varepsilon^{\mu\nu\alpha \beta} R_{\alpha \beta \rho \sigma} R^{\sigma}_{\;\; \nu} \phi^\rho \phi_\mu,\nonumber\\
L_4 &=&  \varepsilon^{\mu\nu\rho\sigma} R_{\rho\sigma \alpha\beta} R^{\alpha \beta}_{\;\;\;\; \mu\nu} \phi^\lambda \phi_\lambda,\nonumber
\eqn
with $\phi^\mu \equiv \nabla^\mu \phi$.
The term $\mathcal{L}_{\rm PV2}$, which contains the second derivatives of the scalar field, is described by
\bqn
\mathcal{L}_{\rm PV2} &=& \sum_{\mathcal{A}=1}^7 b_{\mathcal{A}} (\phi,\phi^\lambda \phi_\lambda) M_{\mathcal{A}},\\
M_1 &=& \varepsilon^{\mu\nu \alpha \beta} R_{\alpha \beta \rho\sigma} \phi^\rho \phi_\mu \phi^\sigma_\nu,\nonumber\\
M_2 &=& \varepsilon^{\mu\nu \alpha \beta} R_{\alpha \beta \rho\sigma} \phi^\rho_\mu \phi^\sigma_\nu, \nonumber\\
M_3 &=& \varepsilon^{\mu\nu \alpha \beta} R_{\alpha \beta \rho\sigma} \phi^\sigma \phi^\rho_\mu \phi^\lambda_\nu \phi_\lambda, \nonumber\\
M_4 &=& \varepsilon^{\mu\nu \alpha \beta} R_{\alpha \beta \rho\sigma} \phi_\nu \phi_\mu^\rho \phi^\sigma_\lambda \phi^\lambda, \nonumber\\
M_5 &=& \varepsilon^{\mu\nu \alpha \beta} R_{\alpha \rho\sigma \lambda } \phi^\rho \phi_\beta \phi^\sigma_\mu \phi^\lambda_\nu, \nonumber\\
M_6 &=& \varepsilon^{\mu\nu \alpha \beta} R_{\beta \gamma} \phi_\alpha \phi^\gamma_\mu \phi^\lambda_\nu \phi^\lambda, \nonumber\\
M_7 &=& (\nabla^2 \phi) L_1,\nonumber
\eqn
with $\phi^{\sigma}_\nu \equiv \nabla^\sigma \nabla_\nu \phi$. 

Variation of the action with respect to the metric tensor $g_{\mu\nu}$,  one obtains the field equation of the theory, which is \cite{Qiao:2021fwi},
\bqn
G_{\mu\nu} + 16 \pi G (C_{\mu\nu} +  A_{\mu\nu} +  B_{\mu\nu} ) = 8 \pi G (T_{\mu\nu}^{\rm mat}+ T_{\mu\nu}^{\phi}), \nb\\\lb{field}
\eqn
where $G_{\mu\nu}=R_{\mu\nu}- \frac{1}{2}g_{\mu\nu} R$ is the Einstein tensor and $T^{\rm mat}_{\mu\nu}$ is the matter stress-energy tensor. The tensor $C_{\mu\nu}$ and the stress-energy tensor $T_{\mu\nu}^{\phi}$ for the scalar field are defined by
\bqn
C_{\mu\nu} 
&\equiv& \;^*R^\beta{}_{\mu\nu}{}^\alpha \nabla_\alpha\nabla_\beta \vartheta + (\nabla_\alpha \vartheta) \varepsilon^{\alpha \beta \gamma}_{\;\;\;\;\;\;\mu} \nabla_\gamma R_{\nu \beta }, 
\eqn
and
\bqn\lb{sft}
T^{\phi}_{\mu\nu}\equiv \nabla_\mu \phi\nabla_\nu \phi - \frac{1}{2}g_{\mu\nu} \nabla^\alpha  \phi\nabla_\alpha \phi + g_{\mu\nu}V(\phi).
\eqn
The expressions of $A_{\mu\nu}$ and $B_{\mu\nu}$ are given in Appendix A. The equation of the scalar field can be obtained by varying the action (\ref{action}) with respect to the scalar field $\phi$, which gives
\bqn
\nabla^2 \phi + \frac{dV(\phi)}{d\phi} + \frac{1}{4}\vartheta_{,\phi} \;^*R R + F_{\phi} =0,\lb{scalar_Eq}
\eqn
where $\vartheta_{,\phi} = d\vartheta/d\phi$, $F_{\phi}$ is given by (\ref{Fphi}) in the appendix A.

\section{Slowly rotating black hole solution}
 \renewcommand{\theequation}{3.\arabic{equation}} \setcounter{equation}{0}
 
In this section we study slowly rotating black holes in the chrial scalar-tensor theory. Note that in this paper, we set $V(\phi)=0$ and $T_{\mu\nu}^{\rm mat}=0$ for simplicity.

 \subsection{Approximation Scheme}

The analytical study of stationary and axisymmetric line elements in the chiral scalar-tensor theory without the aid of any approximation scheme is a challenging task. Instead, one has to employ a couple of approximations to get the analytical solutions. In this section, following the strategies adopted in refs.~\cite{Yagi:2012ya, Cano:2019ore}, we consider the stational ad axisymmetric black hole solutions in the chiral scalar-tensor theory with two approximation schemes, the small-coupling approximation and the slow-rotation approximation. For the small-coupling approximation, we treat the corrections to the GR black hole solution (the Kerr black hole in this paper) from all the coupling terms in the action (\ref{PV_L}) as small deformation of GR. In this way, one can perform the following the expansion on the metric of the black hole solution as
\bqn
g_{\mu\nu} &\equiv& \sum_{n=0} \zeta^n g^{(n)}_{\mu\nu} \nb\\
&=& g^{(0)} _{\mu\nu}+\zeta g^{(1)}_{\mu\nu}+\zeta^2 g^{(2)} _{\mu\nu}+\zeta^3 g^{(3)}_{\mu\nu} + \zeta^4 g^{(4)}_{\mu\nu} + \cdots ,\nb\\
\eqn
where $\zeta$ is a parameter that labels the order of the small-coupling approximation with $\zeta \sim \vartheta, \vartheta_{,\phi}, a_A, a_{A, \phi}, a_{A, X}, b_A, b_{A,\phi}, b_{A, X}$, etc with $X=\phi_\mu \phi^\mu$. Here $g^{(0)}_{\mu\nu}$ is the background metric that satisfies the Einstein equations, such as the Kerr metric, while $g^{(n)}_{\mu\nu}$ denotes the $n$-th small-coupling corrections to the Kerr black hole. 

For the slow-rotation approximation, one treats the dimensionless spin parameter $\chi = a/M$ as a small parameter $\chi <1$ with $M$ being the mass of the black hole. Then we can expand each $g^{(n)}_{\mu\nu}$ in the slow-rotation expansion via
\bqn
g^{(0)}_{\mu \nu} &\equiv & \sum_{m=0} \chi^m g^{(m, 0)}_{\mu\nu} \nb\\
&=& g^{(0,0)}_{\mu \nu} + \chi g^{ (1,0)}_{\mu \nu} + \chi^2 g^{(2,0)}_{\mu \nu} + \chi^3 g^{(3,0)}_{\mu \nu} + \cdots, \\
g^{(1)}_{\mu \nu} &\equiv & \sum_{m=0} \chi^m g^{(m, 1)}_{\mu\nu} \nb\\
&=&   g^{(0,1)}_{\mu \nu} +  \chi g^{(1,1)}_{\mu \nu} + \chi^2 g^{(2,1)}_{\mu \nu} + \chi^3 g^{(3,1)}_{\mu \nu} + \cdots,\\
g^{(2)}_{\mu \nu} &\equiv& \sum_{m=0} \chi^m g^{(m, 2)}_{\mu\nu} \nb\\
&& = g^{(0,2)}_{\mu \nu} + \chi g^{(1,2)}_{\mu \nu} +\chi^2 g^{(2,2)}_{\mu \nu} + \chi^3 g^{(3,2)}_{\mu \nu} + \cdots \\
\cdots &&\cdots \cdots.
\eqn
Here we note that $g_{\mu\nu}^{(m, n)}$ is at the order of $O(\zeta^n \chi^m)$.

One can expand the Kerr metric by treating $\chi$ as a small quantity to obtain $g^{(m, 0)}_{\mu\nu}$. The Kerr metric in the Boyer-Lindquist coordinates $(t, r, \theta, \varphi)$ can be written as
\bqn
ds^2 &=& - \left(1- \frac{2 M r}{\Sigma}\right)dt^2- \frac{4 M a r \sin^2{\theta}}{\Sigma} dt d\varphi \nb\\
&&+ \frac{\Sigma}{\Delta} dr^2 + \Sigma d\theta^2\nb\\
&&+ \left(r^2 + a^2 + \frac{2 M a^2 r \sin^2{\theta}}{\Sigma}\right)\sin^2\theta d\varphi^2,
\eqn
where
\bqn
\Sigma \equiv r^2 + a^2 \cos^2\theta, \\
\Delta \equiv r^2 - 2 M r +a^2,
\eqn
with $M$ being the mass of the black hole and $a=J/M$ with $J$ representing the spin angular momentum of the black hole. The dimensionless spin $\chi$ is defined as $\chi=a/M=J/M^2$. By expanding the above metric to desired order of $\chi^m$, one then gets the expressions of $g^{(0,0)}$, $g^{(1, 0)}$, $\cdots$, $g^{(m, 0)}$. For the solution in the dynamical Chern-Simons gravity \cite{Yagi:2012ya, Cardenas-Avendano:2018ocb}, one has $g_{\mu\nu}^{(m,1)}=0$, $g_{\mu\nu}^{(1,2)}=0$ except $g_{t\varphi}^{(1,2)} \neq 0$. For the contributions of the chiral scalar-tensor theory to the metric, as we analyzed later, they can appear at the order of $O(\chi^2\zeta^3)$ for $g_{t\varphi}^{(2,3)}$ with other components vanishing at the same order. For simplify, in this paper, we only focus on the metric up to the order of $O(\chi^2 \zeta^3)$ and consider the contribution at higher order in a future work.

For the scalar field $\phi$, for small-coupling approximation, we expand it as follows,
\bqn
\phi &=& \sum_{n=0} \zeta^n \phi^{(n)} \nb\\
&=& \phi^{(0)}+ \zeta \phi^{(1)} + \zeta^2 \phi^{(2)} + \zeta^3 \phi^{(3)} + \cdots.
\eqn
Considering the background spacetime, the Kerr metric, is a vaccum solution of GR, the background scalar field $\phi^{(0)}$ vanishes, i.e., $\phi^{(0)}=0$. $\phi^{(3)}$ is at the order of $O(\zeta^3)$, which does not contribute to the metric at order of $O(\chi^2)$, we do not consider in the rest of this paper. For $\phi^{(1)}$ and $\phi^{(2)}$, they can be expanded further in the slow-rotation approximation as
\bqn
\phi^{(1)} &=& \sum_{m=0} \chi^m \phi^{(m,1)} \nb\\
&=& \phi^{(0,1)} + \chi \phi^{(1,1)} + \chi^2 \phi^{(2, 1)} + \chi^3 \phi^{(3,1)}+\cdots, \\
\phi^{(2)} &=& \sum_{m=0} \chi^m \phi^{(m,2)} \nb\\
&=& \phi^{(0,2)} + \chi \phi^{(1,2)} + \chi^2 \phi^{(2, 2)} + \chi^3 \phi^{(3,2)}+\cdots.
\eqn
Considering that the parity-violating effects vanishes for spherical symmetry background, one has $\phi^{(0,1)} =0=\phi^{(0,2)}$. Since $\phi^{(3,1)}$ and $\phi^{(3,2)}$ do not contribute to the metric at the order of $O(\chi^2 \zeta^3)$, we do not consider them in the rest of this paper. Thus, the scalar field is simply expanded as
\bqn
\phi &\simeq&  \zeta \chi \phi^{(1,1)} + \zeta \chi^2 \phi^{(2,1)} \nb\\
&& + \zeta^2 \chi \phi^{(1,2)} + \zeta^2 \chi^2 \phi^{(2,2)}.
\eqn

 \subsection{Black hole solution in dynamical Chern-Simons gravity: up to $\mathcal{O}(\zeta^2 \chi^2)$}

 Let us first give a quick review of the slowly rotating solution in the dynamical Chern-Simons gravity. The slowly rotating solution of the dynamical Chern-Simons gravity have been consideried on a series papers, see refs.~\cite{Yagi:2012ya, Cano:2019ore, Yunes:2009hc, Konno:2009kg, Cardenas-Avendano:2018ocb, Ali-Haimoud:2011zme} for details. 

 In the dynamical Chern-Simons gravity, up to the order $O(\zeta^2 \chi^2)$, the non-vanishing $g^{(m,n)}_{\mu\nu}$ are given by \cite{Yagi:2012ya} (see also \cite{Cardenas-Avendano:2018ocb})
 \bqn
 \zeta^2 \chi^2 g^{(2,2)}_{tt} &=& \frac{\vartheta_{,\phi}^2}{\kappa_g} \chi^2 \frac{M^3}{r^3}\Bigg[\frac { 2 0 1 } { 1 7 9 2 } \Big(1+\frac{M}{r}+\frac{4474}{4221} \frac{M^2}{r^2} \nb\\
 && -\frac{2060}{469} \frac{M^3}{r^3}+\frac{1500}{469} \frac{M^4}{r^4} -\frac{2140}{201} \frac{M^5}{r^5} \nb\\
 && +\frac{9256}{201} \frac{M^6}{r^6}-\frac{5376}{67} \frac{M^7}{r^7}\Big)\left(3 \cos ^2 \theta-1\right) \nb\\
&&  -\frac{5}{384} \frac{M^2}{r^2}\Big(1+100 \frac{M}{r}+194 \frac{M^2}{r^2} \nb\\
&&  +\frac{2220}{7} \frac{M^3}{r^3} -\frac{1512}{5} \frac{M^4}{r^4}\Big)\Bigg], \\
\zeta^2 \chi^2 g^{(2,2)}_{rr} &=& 
\frac{\vartheta_{,\phi}^2}{\kappa_g} \chi^2 \frac{M^3}{r^3 f(r)^2}\Bigg[\frac { 2 0 1 } { 1 7 9 2 } f ( r ) \Big(1+\frac{1459}{603} \frac{M}{r} \nb\\
&& +\frac{20000}{4221} \frac{M^2}{r^2}+\frac{51580}{1407} \frac{M^3}{r^3} -\frac{7580}{201} \frac{M^4}{r^4} \nb\\
&& -\frac{22492}{201} \frac{M^5}{r^5}-\frac{40320}{67} \frac{M^6}{r^6}\Big)\left(3 \cos ^2 \theta-1\right)\nb \\
&& -\frac{25}{384} \frac{M}{r}\Big(1+3 \frac{M}{r}+\frac{322}{5} \frac{M^2}{r^2}+\frac{198}{5} \frac{M^3}{r^3} \nb\\
&& +\frac{6276}{175} \frac{M^4}{r^4}-\frac{17496}{25} \frac{M^5}{r^5}\Big)\Bigg] , \\
\zeta^2 \chi^2 g^{(2,2)}_{\theta \theta} &=& \frac{\vartheta_{,\phi}^2}{\kappa_g} \chi^2 \frac{201M^3}{1792r}\Big(1+\frac{1420}{603} \frac{M}{r} +\frac{18908}{4221} \frac{M^2}{r^2} \nb\\
&& +\frac{1480}{603} \frac{M^3}{r^3} +\frac{22460}{1407} \frac{M^4}{r^4} +\frac{3848}{201} \frac{M^5}{r^5}\nb\\
&& +\frac{5376}{67} \frac{M^6}{r^6}\Big)\left(3 \cos ^2 \theta-1\right), \\
\zeta^2 \chi^2 g^{(2,2)}_{\varphi \varphi} &=& \zeta^2 \chi^2  g^{(2,2)}_{\theta \theta} \sin^2\theta ,\\
\zeta^2 \chi g^{(1,2)}_{t \varphi} &=& \frac{5}{8} \frac{\vartheta_{,\phi}^2}{\kappa_g} M \chi \frac{M^4}{r^4}\left(1+\frac{12}{7} \frac{M}{r}+\frac{27}{10} \frac{M^2}{r^2}\right) \sin ^2 \theta,\nb\\
 \eqn
where $f(r)\equiv 1- 2 M/r$. For other parts of the solution, one has
\bqn
g^{(0,1)}_{\mu\nu}=0, \;\;\; g^{(1,1)}_{\mu\nu}=0, \\
g^{(2,1)}_{\mu\nu}=0,\;\;\;  g^{(0,2)}_{\mu\nu}=0, 
\eqn
and
\bqn 
g^{(1,2)}_{tt}=g^{(1,2)}_{rr}=g^{(1,2)}_{\theta\theta} =g^{(1,2)}_{\varphi\varphi}=0,\\
g^{(2,2)}_{t\varphi}=0.
\eqn
This solution has been extended to $O(\zeta^2\chi^5)$ in \cite{Maselli:2017kic} and $O(\zeta^2\chi^{21})$ in \cite{Cano:2019ore}. For our purpose of calculating the contributions to the solution from the new terms in the chiral scalar-tensor theory beyond Chern-Simons term up to $O(\zeta^3 \chi^2)$, we do not write these high-order contributions here and refer the reader to refs. \cite{Maselli:2017kic, Cano:2019ore, Cardenas-Avendano:2018ocb}.

The solution for the scalar field gives
\bqn
\zeta \chi \phi^{(1,1)} &=& \frac{5}{8} \vartheta_{,\phi} \chi \frac{\cos\theta}{r^2} \left(1+ \frac{2M}{r} + \frac{18 M^2}{r^2}\right).
\eqn
For other parts, one has $\zeta \chi^2 \phi^{(2,1)} = 0$, $\zeta^2 \chi \phi^{(1,1)} =0$, and $\zeta^2 \chi^2 \phi^{(1,1)} =0$.

 \subsection{Black hole solution in chiral scalar-tensor theory: up to $\mathcal{O}(\zeta^3\chi^2)$}

 Now let us turn to consider the slowly rotating black hole solution in chiral scalar-tensor theory, focusing on the contributions from the new terms in (\ref{PV_L}) beyond the dynamical Chern-Simons gravity.  

 Since the terms of $O(\zeta^3 \chi^2)$ are parity odd, we only need to consider the odd parity sector of the metric, $g_{t \varphi}^{(2,3)}$, while the even sector of the metric at the order of $O(\zeta^3 \chi^2)$ simply vanish, i.e., $g_{tt}^{(2,3)} = g_{rr}^{(2,3)}=g_{\theta\theta}^{(2,3)}=g_{\varphi\varphi}^{(2,3)}=0$. With this fact, to get the solution at the order of $O(\zeta^3 \chi^2)$, one can parametrize $g_{t \varphi}^{(2,3)}$ in the form of
 \bqn
 g^{(2,3)}_{t\varphi} = \omega(r, \theta) r^2 \sin^2\theta,
 \eqn
 where $\omega(r, \theta)$ is a function of $r$ and $\theta$. Then substituting it into the field equations (\ref{field}) and expand them up to the order of $O(\zeta^3 \chi^2)$, one gets
 \begin{widetext}
 \bqn
 &&r(r-2M)\partial^2_rw +4(r-2M) \partial_r w + \partial^2_\theta w + 3 \cot\theta \partial_\theta w \nb\\
 &&~~~~~=\frac{a_1\vartheta_{,\phi}^2}{\kappa_g}(r-2M) \cos \theta \frac{6525M}{8r^{11}}  \left(1+\frac{340 M}{87 r}+\frac{1628 M^2}{145 r^2} -\frac{2468 M^3}{145 r^3}-\frac{9228 M^4}{145 r^4}-\frac{119232 M^5}{725 r^5}\right)\nb\\
 &&~~~~~~~~ + \frac{a_2 \vartheta_{,\phi}^2}{\kappa_g} (r-2M)\cos \theta \frac{1575M}{8r^{11}}  \left(1+\frac{32 M}{7 r}+\frac{516 M^2}{35 r^2}  -\frac{96 M^3}{7 r^3}-\frac{396 M^4}{5 r^4}-\frac{41472 M^5}{175 r^5}\right) \nb\\
&&~~~~~~~~ + \frac{a_3\vartheta_{,\phi}^2}{\kappa_g} (r-2M)\cos\theta \frac{2475 M}{8r^{11}}\left(1+\frac{122 M}{33 r}+\frac{556 M^2}{55 r^2}-\frac{994 M^3}{55 r^3}-\frac{3228 M^4}{55 r^4}-\frac{7776 M^5}{55 r^5} \right)\nb\\
 &&~~~~~~~~ + \frac{a_4 \vartheta_{,\phi}^2}{\kappa_g} (r-2M)\cos \theta \frac{1575M}{2r^{11}}  \left(1+\frac{32 M}{7 r}+\frac{516 M^2}{35 r^2}  -\frac{96 M^3}{7 r^3}-\frac{396 M^4}{5 r^4}-\frac{41472 M^5}{175 r^5}\right) \nb\\
 &&~~~~~~~~ + \frac{b_2\vartheta_{,\phi}^2}{\kappa_g} (r-2M)\cos\theta \frac{13050M}{32 r^{11}} \left(1+\frac{340 M}{87 r}+\frac{1628 M^2}{145 r^2}-\frac{2468 M^3}{145 r^3}-\frac{9228 M^4}{145 r^4}-\frac{119232 M^5}{725 r^5}\right). \lb{equationE}
 \eqn
  \end{widetext}

Following refs.~\cite{Konno:2009kg}, we expand $\omega(r, \theta)$ as
\bqn
\omega(r, \theta) = \sum_{n=1}^{+\infty} \tilde \omega_n(r)\frac{1}{\sin\theta}\partial_\theta P_n(\cos\theta),
\eqn
where $P_n(\cos\theta)$ is the Legendre function of the first kind. From the equation (\ref{equationE}), one observes that the left-hand side of (\ref{equationE}) should be proportional to $\cos\theta$, which indicates 
\bqn
&&r(r-2M)\partial^2_rw +4(r-2M) \partial_r w + \partial^2_\theta w + 3 \cot\theta \partial_\theta w \nb\\
&&~~~~~~~~~~~ ~~~~~ \propto \cos \theta.
\eqn
This property requires
\bqn
\omega(r,\theta) = -{\tilde \omega}_1(r) \cos\theta,
\eqn
with $\tilde \omega_n(r)=0$ for $n >1$. Substituting the above relation into Eq.~(\ref{equationE}) and then solving for $\tilde \omega_1(r)$, one gets
\begin{widetext}
 \bqn
 \tilde \omega_1&=& \frac{a_1 \vartheta_{,\phi}^2}{\kappa_g} \left[\frac{3726 M^6}{5 r^{15}}+\frac{504171 M^5}{1540 r^{14}}+\frac{14787 M^4}{154 r^{13}} -\frac{16377 M^3}{176 r^{12}}-\frac{75843 M^2}{1936 r^{11}}-\frac{419739 M}{33880  r^{10}} \right. \nb\\
 &&~~~~~~~ \left. +\frac{2103}{630784 M^9} \left(1-\frac{M}{r}-\frac{2 M^2}{3 r^2}-\frac{2 M^3}{3 r^3} -\frac{4 M^4}{5 r^4}-\frac{16 M^5}{15 r^5}-\frac{32 M^6}{21 r^6} -\frac{16 M^7}{7 r^7}-\frac{32 M^8}{9 r^8}-\frac{256 M^9}{45 r^9}\right) \right.\nb\\
 &&~~~~~~~~ \left. + \frac{2103 (r-2M)}{1261568 M^{10}}\ln \frac{r-2M}{r} \right] \nb\\
&&+ \frac{a_2 \vartheta_{,\phi}^2}{\kappa_g} \left[-\frac{51057 M}{16940 r^{10}}-\frac{21393 M^2}{1936 r^{11}}-\frac{5157 M^3}{176 r^{12}}+\frac{5319 M^4}{308 r^{13}}+\frac{150741 M^5}{1540 r^{14}}+\frac{1296 M^6}{5r^{15}}\right. \nb\\
 &&~~~~~~~ \left. +\frac{2103}{630784 M^9} \left(1-\frac{M}{r}-\frac{2 M^2}{3 r^2}-\frac{2 M^3}{3 r^3} -\frac{4 M^4}{5 r^4}-\frac{16 M^5}{15 r^5}-\frac{32 M^6}{21 r^6} -\frac{16 M^7}{7 r^7}-\frac{32 M^8}{9 r^8}-\frac{256 M^9}{45 r^9}\right) \right.\nb\\
 &&~~~~~~~~ \left. + \frac{2103 (r-2M)}{1261568 M^{10}}\ln \frac{r-2M}{r} \right] \nb\\
 && -\frac{a_3 \vartheta_{,\phi}^2}{\kappa_g} \frac{75M}{16 r^{10}}\left[1+\frac{3 M}{r}+\frac{34 M^2}{5 r^2}-\frac{42 M^3}{5 r^3}-\frac{612 M^4}{25 r^4}-\frac{1296 M^5}{25 r^5} \right] \nb\\
 &&+ \frac{4 a_4 \vartheta_{,\phi}^2}{\kappa_g} \left[-\frac{51057 M}{16940 r^{10}}-\frac{21393 M^2}{1936 r^{11}}-\frac{5157 M^3}{176 r^{12}}+\frac{5319 M^4}{308 r^{13}}+\frac{150741 M^5}{1540 r^{14}}+\frac{1296 M^6}{5r^{15}}\right. \nb\\
 &&~~~~~~~ \left. +\frac{2103}{630784 M^9} \left(1-\frac{M}{r}-\frac{2 M^2}{3 r^2}-\frac{2 M^3}{3 r^3} -\frac{4 M^4}{5 r^4}-\frac{16 M^5}{15 r^5}-\frac{32 M^6}{21 r^6} -\frac{16 M^7}{7 r^7}-\frac{32 M^8}{9 r^8}-\frac{256 M^9}{45 r^9}\right) \right.\nb\\
 &&~~~~~~~~ \left. + \frac{2103 (r-2M)}{1261568 M^{10}}\ln \frac{r-2M}{r} \right] \nb\\
 &&+ \frac{b_2 \vartheta_{,\phi}^2}{\kappa_g} \left[-\frac{419739 M}{67760 r^{10}}-\frac{75843 M^2}{3872 r^{11}}-\frac{16377 M^3}{352 r^{12}}+\frac{14787 M^4}{308 r^{13}}+\frac{504171 M^5}{3080 r^{14}}+\frac{1863 M^6}{5 r^{15}}\right. \nb\\
 &&~~~~~~~ \left. +\frac{2103}{1261568 M^9} \left(1-\frac{M}{r}-\frac{2 M^2}{3 r^2}-\frac{2 M^3}{3 r^3} -\frac{4 M^4}{5 r^4}-\frac{16 M^5}{15 r^5}-\frac{32 M^6}{21 r^6} -\frac{16 M^7}{7 r^7}-\frac{32 M^8}{9 r^8}-\frac{256 M^9}{45 r^9}\right) \right.\nb\\
 &&~~~~~~~~ \left. + \frac{2103 (r-2M)}{2523136 M^{10}}\ln \frac{r-2M}{r} \right], 
 \eqn
 which leads to
 \bqn
 \zeta^3 \chi^2 g^{(2,3)}_{t\varphi} &= & - \frac{a_1 \vartheta_{,\phi}^2}{\kappa_g} \chi^2 r^2 \left[\frac{3726 M^6}{5 r^{15}}+\frac{504171 M^5}{1540 r^{14}}+\frac{14787 M^4}{154 r^{13}} -\frac{16377 M^3}{176 r^{12}}-\frac{75843 M^2}{1936 r^{11}}-\frac{419739 M}{33880  r^{10}} \right. \nb\\
 &&~~~~~~~ \left. +\frac{2103}{630784 M^9} \left(1-\frac{M}{r}-\frac{2 M^2}{3 r^2}-\frac{2 M^3}{3 r^3} -\frac{4 M^4}{5 r^4}-\frac{16 M^5}{15 r^5}-\frac{32 M^6}{21 r^6} -\frac{16 M^7}{7 r^7}-\frac{32 M^8}{9 r^8}-\frac{256 M^9}{45 r^9}\right) \right.\nb\\
 &&~~~~~~~~ \left. + \frac{2103 (r-2M)}{1261568 M^{10}}\ln \frac{r-2M}{r} \right] \cos\theta \sin^2\theta \nb\\
&& - \frac{a_2 \vartheta_{,\phi}^2}{\kappa_g} \chi^2 r^2\left[-\frac{51057 M}{16940 r^{10}}-\frac{21393 M^2}{1936 r^{11}}-\frac{5157 M^3}{176 r^{12}}+\frac{5319 M^4}{308 r^{13}}+\frac{150741 M^5}{1540 r^{14}}+\frac{1296 M^6}{5r^{15}}\right. \nb\\
 &&~~~~~~~ \left. +\frac{2103}{630784 M^9} \left(1-\frac{M}{r}-\frac{2 M^2}{3 r^2}-\frac{2 M^3}{3 r^3} -\frac{4 M^4}{5 r^4}-\frac{16 M^5}{15 r^5}-\frac{32 M^6}{21 r^6} -\frac{16 M^7}{7 r^7}-\frac{32 M^8}{9 r^8}-\frac{256 M^9}{45 r^9}\right) \right.\nb\\
 &&~~~~~~~~ \left. + \frac{2103 (r-2M)}{1261568 M^{10}}\ln \frac{r-2M}{r} \right]\cos\theta  \sin^2\theta \nb\\
 && +\frac{a_3 \vartheta_{,\phi}^2}{\kappa_g} \chi^2 \frac{75M}{16 r^{8}}\left[1+\frac{3 M}{r}+\frac{34 M^2}{5 r^2}-\frac{42 M^3}{5 r^3}-\frac{612 M^4}{25 r^4}-\frac{1296 M^5}{25 r^5} \right] \cos\theta  \sin^2\theta\nb\\
 &&- \frac{4 a_4 \vartheta_{,\phi}^2}{\kappa_g} \chi^2 r^2\left[-\frac{51057 M}{16940 r^{10}}-\frac{21393 M^2}{1936 r^{11}}-\frac{5157 M^3}{176 r^{12}}+\frac{5319 M^4}{308 r^{13}}+\frac{150741 M^5}{1540 r^{14}}+\frac{1296 M^6}{5r^{15}}\right. \nb\\
 &&~~~~~~~ \left. +\frac{2103}{630784 M^9} \left(1-\frac{M}{r}-\frac{2 M^2}{3 r^2}-\frac{2 M^3}{3 r^3} -\frac{4 M^4}{5 r^4}-\frac{16 M^5}{15 r^5}-\frac{32 M^6}{21 r^6} -\frac{16 M^7}{7 r^7}-\frac{32 M^8}{9 r^8}-\frac{256 M^9}{45 r^9}\right) \right.\nb\\
 &&~~~~~~~~ \left. + \frac{2103 (r-2M)}{1261568 M^{10}}\ln \frac{r-2M}{r} \right]\cos\theta  \sin^2\theta\nb\\
 &&+ \frac{b_2 \vartheta_{,\phi}^2}{\kappa_g} \left[-\frac{419739 M}{67760 r^{10}}-\frac{75843 M^2}{3872 r^{11}}-\frac{16377 M^3}{352 r^{12}}+\frac{14787 M^4}{308 r^{13}}+\frac{504171 M^5}{3080 r^{14}}+\frac{1863 M^6}{5 r^{15}}\right. \nb\\
 &&~~~~~~~ \left. +\frac{2103}{1261568 M^9} \left(1-\frac{M}{r}-\frac{2 M^2}{3 r^2}-\frac{2 M^3}{3 r^3} -\frac{4 M^4}{5 r^4}-\frac{16 M^5}{15 r^5}-\frac{32 M^6}{21 r^6} -\frac{16 M^7}{7 r^7}-\frac{32 M^8}{9 r^8}-\frac{256 M^9}{45 r^9}\right) \right.\nb\\
 &&~~~~~~~~ \left. + \frac{2103 (r-2M)}{2523136 M^{10}}\ln \frac{r-2M}{r} \right]\cos\theta  \sin^2\theta. \lb{ct_sol}
 \eqn
 \end{widetext}
The contributions from the new terms in the chiral scalar-tensor theory are found to be coupled to the Chern-Simons term. At the order of $O(\zeta^3\chi^2)$, the new terms in the chiral scalar-tensor theory do not produce effects independently but instead modify the solution through their coupling to the Chern-Simons contributions. 

In the next section, we analyze the main properties of this new solution, focusing on its behavior in the weak-field limit as well as the characteristics of its ergosphere and horizon.  

\subsection{Validity of the approximate solution}

\textcolor{black}{In this subsection, we discuss the validity of the approximate solution presented above.  }

\textcolor{black}{To construct the solution, we employed perturbative techniques based on small parameter expansions in terms of the spin parameter \(\chi\) and the coupling constant \(\zeta\). Specifically, we truncated the perturbative series at the order \(O(\zeta^3 \chi^2)\). This approach follows the well-established framework used in dynamical Chern-Simons (dCS) gravity \cite{Yagi:2012ya, Cano:2019ore, Yunes:2009hc, Konno:2009kg, Cardenas-Avendano:2018ocb, Ali-Haimoud:2011zme, Maselli:2017kic} and scalar-Gauss-Bonnet (sGB) gravity \cite{Maselli:2015tta}. In these theories, truncation at specific orders in \(\zeta\) and \(\chi\) has been rigorously validated.  }

\textcolor{black}{For example, in dCS gravity, the slowly rotating black hole solution has been obtained to orders of \(O(\zeta^2 \chi^5)\) in \cite{Maselli:2017kic} and \(O(\zeta^2 \chi^{21})\) in \cite{Cano:2019ore}. It has been shown that the expansion up to \(O(\chi^{14})\) remains valid for \(\chi < 0.7\), with relative errors in the expansion coefficients below \(0.81\%\) \cite{Cano:2019ore}. Similarly, in sGB gravity, slowly rotating black hole solutions have been derived at orders of \(O(\zeta^2 \chi^{21})\) in \cite{Cano:2019ore} and \(O(\zeta^7 \chi^5)\) in \cite{Maselli:2015tta}. In particular, the solution at \(O(\zeta^7 \chi^2)\) has been shown to agree excellently with the exact result in \(\zeta\), albeit approximate in \(\chi^2\), with discrepancies below \(1\%\) for \(\zeta \sim 0.6\) \cite{Maselli:2015tta}. These results confirm that the leading-order terms dominate in the weak-coupling regime (\(\zeta \ll 1\)), thereby justifying our truncation at \(O(\zeta^3 \chi^2)\).  }

\textcolor{black}{The chiral scalar-tensor theory shares a similar mathematical structure with dCS and sGB gravity. Specifically, the chiral scalar-tensor theory extends dCS gravity by incorporating couplings between the first and second derivatives of the scalar field and spacetime curvature invariants. At leading order in the coupling constants, the parity-violating terms contribute to the spacetime metric in analogous ways, albeit at different orders. The iterative procedure for solving the field equations follows the same hierarchy of \(\zeta^n \chi^m\). Given the proven convergence of this hierarchy in dCS gravity \cite{Cano:2019ore} and sGB gravity \cite{Cano:2019ore, Maselli:2015tta}, we expect similar convergence behavior in the chiral scalar-tensor theory, as long as \(\zeta\) and \(\chi\) remain within the validated ranges.  }

\textcolor{black}{The approximate solution is, of course, only valid in the regime where \(\chi \ll 1\) and \(\zeta \ll 1\). However, the precise upper limits of \(\chi\) and \(\zeta\) for which the approximation remains valid are not yet known. The most reliable way to determine these limits would be to compare the \(O(\zeta^3 \chi^2)\)-accurate metric with a fully numerical solution. Unfortunately, such a numerical solution is currently unavailable for the chiral scalar-tensor theory. Instead, following the approach in \cite{Ayzenberg:2014aka}, we can estimate the error introduced by truncating the perturbative series by considering the next-order terms expected in the \(\chi \ll 1\) and \(\zeta \ll 1\) expansions.  }

\textcolor{black}{To estimate the validity range of \(\chi\), we use the Kerr solution at \(O(\zeta^0 \chi^2)\) as a benchmark, since all terms at orders \(O(\zeta^1 \chi^2)\), \(O(\zeta^2 \chi^2)\), and \(O(\zeta^3 \chi^2)\) are expected to be smaller than the \(O(\zeta^0 \chi^2)\) term. Based on the analysis in \cite{Ayzenberg:2014aka}, requiring a relative error of \(\sim 1\%\), we find:  
\begin{align}
    \chi \lesssim 0.01^{1/3} \quad &\text{(spatial-time components)}, \\
    \chi \lesssim 0.01^{1/4} \quad &\text{(diagonal components)}.  
\end{align}  
These conditions correspond to \(\chi \lesssim 0.21\) and \(\chi \lesssim 0.31\), respectively. To be conservative, we adopt \(\chi \lesssim 0.2\) as a rough validity range for the spin parameter, with a relative error of approximately \(1\%\).  }

\textcolor{black}{For the coupling constant \(\zeta\), we note that all terms at orders \(O(\zeta^1 \chi^2)\), \(O(\zeta^2 \chi^2)\), and \(O(\zeta^3 \chi^2)\) are smaller than the \(O(\zeta^0 \chi^2)\) term, providing that $\zeta \chi \ll 1$ with $\zeta \lesssim 1$. Thus, we expect the approximate solution to remain valid as long as \(\zeta \lesssim 1\), ensuring the convergence of the expansion.  }

\section{Properties of black hole solution}
 \renewcommand{\theequation}{4.\arabic{equation}} \setcounter{equation}{0}

 In this section, we explore the properties of the slowly rotating solution obtained in the above section. Specifically, we check the black hole solution in the weak field limit and its ergosphere and horizion, respectively. 

 \subsection{Weak field limit}

  In the weak field limit, the the componnent $g_{t\varphi}^{(2,3)}$ of the metric can be expanded in  terms of $1/r$ as
 \bqn
 \zeta^3 \chi^2 g^{(2,3)}_{t\varphi} &\simeq& \frac{a_1 \vartheta_{,\phi}^2}{\kappa_g} \chi^2 \frac{2175 M}{176r^{8}} \cos\theta \sin^2\theta \nb\\
 && + \frac{a_2 \vartheta_{,\phi}^2}{\kappa_g} \chi^2 \frac{525 M}{176r^{8}} \cos\theta \sin^2\theta \nb\\
 &&+ \frac{a_3 \vartheta_{,\phi}^2}{\kappa_g} \chi^2 \frac{75 M}{16r^{8}} \cos\theta \sin^2\theta \nb\\
 &&+  \frac{a_4 \vartheta_{,\phi}^2}{\kappa_g} \chi^2 \frac{525 M}{44r^{8}} \cos\theta \sin^2\theta \nb\\
 && + \frac{b_2 \vartheta_{,\phi}^2}{\kappa_g} \chi^2 \frac{2175 M}{88r^{8}} \cos\theta \sin^2\theta  +\mathcal{O}\left(\frac{1}{r^9}\right).\nb\\
 \eqn
It is evident that these new contributions (as presented in Eq.~(\ref{ct_sol})) are highly suppressed in the far field limit, decaying as $r^{-8}$, which suggests that their effects can only be significant in the strong field regime. 

\subsection{Ergosphere and horizon}

The ergosphere of the black hole is determined by solving the equation  
\bqn
g_{tt} = 0
\eqn
for $r$. Up to the order of $O(\zeta^3 \chi^2)$, the chiral scalar-tensor theory does not introduce any corrections to the $g_{tt}$ component. Including the corrections from Chern-Simons gravity, the ergosphere radius is given by \cite{Yagi:2012ya}:  
\bqn
r_{\rm E} &\simeq &M + M\sqrt{1 - \chi^2 \cos^2\theta}  \nb\\
&& - \frac{915}{28672} \frac{\vartheta_{,\phi}^2}{\kappa_g} M \chi^2 \left(1 + \frac{2816}{915}\sin^2\theta\right).
\eqn
This result is identical to that found in Chern-Simons gravity \cite{Yagi:2012ya}, indicating that the new terms in the chiral scalar-tensor theory do not affect the location of the ergosphere at this order.  

The black hole horizon is determined by solving the equation  
\bqn
g_{tt} g_{\varphi\varphi} - g_{t\varphi}^2 = 0
\eqn
for $r$. Up to the same order, the horizon radius is given by \cite{Yagi:2012ya}:  
\bqn
r_{\rm H} &\simeq& M + \sqrt{M^2 - M^2\chi^2}  \nb\\
&&- \frac{915}{28672} \frac{\vartheta_{,\phi}^2}{\kappa_g} M \chi^2.
\eqn
Similar to the ergosphere, the horizon radius is not modified by the new terms in the chiral scalar-tensor theory up to the order $O(\zeta^3 \chi^2)$.  

In summary, the contributions from the chiral scalar-tensor theory do not introduce any corrections to the locations of the black hole's ergosphere or horizon up to $O(\zeta^3 \chi^2)$. Thus, these locations remain unchanged from those derived in Chern-Simons gravity, highlighting the limited impact of the new terms from the chiral scalar-tensor theory at this perturbative order.

\section{conclusion and outlook}
\renewcommand{\theequation}{5.\arabic{equation}} \setcounter{equation}{0}

The chiral scalar-tensor theory extends Chern-Simons modified gravity by introducing couplings between the first and second derivatives of the scalar field and parity-violating spacetime curvatures. A key feature of this theory is its explicit breaking of parity symmetry in the gravitational sector. In this study, we investigate the effects of parity violation on rotating black hole solutions within the framework of chiral scalar-tensor gravity. Deriving such solutions for parity-violating theories, especially those involving higher derivatives, is a highly challenging task. To address this, we employ two approximation schemes: the small coupling approximation and the slow rotation approximation. These approximations allow us to treat the corrections to the Kerr black hole from all coupling terms in the action as small deformations of general relativity (GR), with the rotation parameter $\chi$ assumed to be a small quantity.  By perturbatively solving the field equations, we find that the contributions of the chiral scalar-tensor theory to the metric appear at quadratic order in spin and cubic order in the coupling constants. At this order, the corrections only affect the time-spatial component of the metric. 

We also briefly explore the properties of this new solution. In the weak-field limit, the contributions to the metric from the chiral scalar-tensor theory are highly suppressed, decaying as $r^{-8}$. This indicates that the effects of the theory become significant only in the strong-field regime, such as near black hole horizons. Additionally, we examine the ergosphere and horizon structure of the solution and find that, up to $O(\zeta^3 \chi^2)$, their locations remain unchanged compared to those in the dynamical Chern-Simons gravity. This suggests that the dominant effects of the chiral scalar-tensor theory on the horizon and ergosphere of the black hole can only be possible in the order higher than $O(\zeta^3 \chi^2)$, warranting further investigation.

There are several possible avenues for extending this study. Firstly, to analytically solve the field equations, we adopt the small coupling approximation and the slow rotation approximation. It is interesting to explore the fast rotating black hole solution without these approximation schemes, which requires numerical solution of the field equations of the chiral scalar-tensor theory. Secondly, the predictions of parity-violating gravity theories, including the chiral scalar-tensor theory, in the strong-field regime, can be used to test or constrain this new solution. One could study how the electromagnetic radiation from accretion disks around a central black hole, including the black hole shadows, quasi-periodic oscillations, geodetic precessions and strong lensing, are modified if the central supermassive black hole is described by the new solution found in this paper. Similarly, gravitational wave observations from extreme mass ratio inspirals (EMRIs) detected by future space-based detectors like LISA (Laser Interferometer Space Antenna) will offer unparalleled precision in probing the near-horizon geometry of rotating black holes. Since EMRIs involve a compact object spiraling into a supermassive black hole, the emitted gravitational waves carry detailed information about the spacetime geometry, including possible parity-violating effects. 

In conclusion, while our study provides an initial exploration of parity-violating effects of the chiral scalar-tensor theory in the rotating black holes, future observational data from black hole imaging, gravitational wave astronomy, etc.,  will be crucial in testing and constraining chiral scalar-tensor gravity and other parity-violating extensions of GR. These efforts will help bridge the gap between theoretical predictions and astrophysical observations, paving the way for a deeper understanding of parity symmetry of gravity in the strong-field regime.

 \section*{Acknowledgments}
 
This work is supported by the National Natural Science Foundation of China under Grants No.~12275238 and No.~11675143, the Zhejiang Provincial Natural Science Foundation of China under Grants No.~LR21A050001 and No.~LY20A050002, the National Key Research and Development Program of China under Grant No. 2020YFC2201503, and the Fundamental Research Funds for the Provincial Universities of Zhejiang in China under Grant No.~RF-A2019015.

\appendix

\section{The expressions of $A_{\mu\nu}$, $B_{\mu\nu}$, and $F_{\phi}$}
\renewcommand{\theequation}{A.\arabic{equation}} \setcounter{equation}{0}

In this Appendix, we list the explicit expression of $A_{\mu\nu}$, $B_{\mu\nu}$ and $F_{,\phi}$ in Eqs.(\ref{field}) and (\ref{scalar_Eq}). These expressions can also be found in \cite{Qiao:2021fwi}. The expression of $A_{\mu\nu}$ in Eq.(\ref{field}) contains four terms,
\bqn
A_{\mu\nu} = \sum_{A=1}^{4} A_{\mu\nu}^{(A)},
\eqn
where
\bqn
A^{(1)}_{\mu\nu}&=& \varepsilon^{\kappa\gamma\alpha \beta} R_{\alpha \beta \rho \sigma} R_{\kappa\gamma}{}^{\rho}{}_{ \lambda} \phi^\sigma \phi^\lambda a_{1,X}\phi_{(\nu}\phi_{\mu)} \nb\\
&& - a_1 \varepsilon^{\lambda\gamma\alpha\beta} \phi_{\sigma}\phi_{(\mu} R^{\sigma}{}_{\rho\alpha\beta}R^{\rho}{}_{\nu)\lambda\gamma} \nb \\
&& + 2 \varepsilon^{\sigma\beta\alpha}{}_{(\nu} \nabla_{\rho} \nabla_{\alpha} \left(a_1 \phi_{\mu)}\phi_{\lambda} \right)   R_{\sigma\beta}{}^{\rho\lambda} \nb\\
&& + 2 \varepsilon^{\sigma\beta\alpha}{}_{(\nu} \nabla_{\alpha} \left(a_1 \phi_{\mu)}\phi_{\lambda} \right) \nabla_{\rho} R_{\sigma\beta}{}^{\rho\lambda} \nb\\
&& + 2 \varepsilon^{\sigma\beta\alpha}{}_{(\nu} \nabla_{\rho} \nabla_{\alpha} \left( a_1\phi_{\lambda}\phi^{\rho}  \right) R^{\lambda}{}_{\mu)\sigma\beta} \nb\\
&& + 2 \varepsilon^{\sigma\beta\alpha}{}_{(\nu} \nabla_{\alpha} \left( a_1\phi_{\lambda}\phi^{\rho} \right) \nabla_{\rho}  R^{\lambda}{}_{\mu)\sigma\beta},\\
A_{\mu\nu}^{(2)} &=&  -\varepsilon^{\delta\tau\alpha \beta} a_{2,X} \phi_{(\mu} \phi_{\nu)} \phi_\tau \phi_\kappa R^{\sigma}{}_{\rho \alpha \beta } R^{\rho }{}_{\sigma \delta \lambda } g^{\lambda\kappa} \nb\\
&& -\varepsilon^{\lambda\kappa\alpha \beta} \phi_\kappa \phi_{(\nu} a_2 R^{\sigma}{}_{\rho \alpha \beta } R^{\rho }{}_{\sigma \lambda \mu) } \nb\\
&& +2\varepsilon^{\sigma \beta\alpha}{}_{ (\nu}  g^{\lambda\kappa}  \nabla_\rho \nabla_\alpha(\phi_\beta \phi_\kappa a_2 R^{\rho }{}_{\mu)\sigma \lambda }) \nb\\
&& +\varepsilon^{\rho\lambda\alpha \beta}  \nabla_\sigma \nabla_\rho (\phi_\lambda \phi_{(\nu} a_2 R^{\sigma}{}_{\mu)\alpha \beta })  \nb\\
&& -\varepsilon_{(\mu}{}^{\rho\alpha \beta}  g^{\lambda\kappa}  \nabla_\sigma \nabla_\lambda (\phi_\rho \phi_\kappa a_2 R^{\sigma}{}_{\nu) \alpha \beta }), \\
A_{\mu\nu}^{(3)} &=& a_{3,X}  \varepsilon^{\delta\tau\alpha \beta} R_{\alpha \beta \rho \sigma} R^{\sigma}{}_{\tau} \phi^\rho \phi_\delta   \phi_{(\nu}\phi_{\mu)} \nb\\
&& + a_3 \varepsilon^{\sigma\omega\alpha\beta} \phi_\rho \phi_\sigma R^{\rho }{}_{(\mu \alpha \beta } R_{\nu)\omega} \nb\\
&& -  \varepsilon^{\rho\beta\alpha}{}_{(\nu} \nabla_\sigma\nabla_\alpha  \left(a_3 \phi_{\mu)} \phi_\rho R^{\sigma}{}_{\beta} \right)  \nb\\
&&  -  \varepsilon^{\sigma\rho\alpha\beta}\nabla_\beta\nabla_\alpha \left(a_3 \phi_{(\nu} \phi_\sigma R_{\mu)\rho} \right) \nb\\
 &&   + \varepsilon^{\sigma\beta\alpha}{}_{(\nu}\nabla_\rho \nabla_\alpha \left(a_3  \phi^\rho \phi_\sigma R_{\mu)}{}_{\beta} \right)  \nb\\
&&- \frac{1}{2}  \varepsilon^{\lambda}{}_{(\nu}{}^{\alpha\beta} \nabla_\sigma \nabla_{\mu)} \left( a_3 \phi_\rho \phi_\lambda R^{\rho\sigma }{}_{ \alpha \beta }  \right) \nb\\
 &&  - \frac{1}{2}  \varepsilon^{\sigma\lambda\alpha\beta} \nabla_\lambda \nabla_{(\nu} \left( a_3 \phi_\rho \phi_\sigma  R^{\rho }{}_{\mu) \alpha \beta } \right) \nb\\
 && + \frac{1}{2}  \varepsilon^{\sigma}{}_{(\nu}{}^{\alpha\beta} \nabla_\lambda\nabla^\lambda \left( a_3 \phi_\rho \phi_\sigma  R^{\rho }{}_{\mu) \alpha \beta }\right) \nb\\
 && + \frac{1}{2}  \varepsilon^{\tau \lambda\alpha\beta}\nabla_\sigma \nabla_\lambda \left( a_3  \phi_\rho \phi_\tau R^{\rho\sigma }{}_{ \alpha \beta } g_{(\nu\mu)} \right) ,\\
A_{\mu\nu}^{(4)} &=& a_{4,X} \varepsilon^{\delta\tau\rho\sigma} R_{\rho\sigma \alpha\beta} R^{\alpha \beta}{}_{ \delta\tau} \phi^\lambda \phi_\lambda \phi_{(\nu}\phi_{\mu)} \nb\\
&& + a_4  \phi_{(\mu} \phi_{\nu)} \varepsilon^{\lambda\kappa\rho\sigma} R_{\rho \sigma \alpha \beta} R^{\alpha \beta }{}_{\lambda\kappa} \nb\\
&& +4 \varepsilon^{\alpha\rho\sigma}{}_{(\nu} \nabla_\alpha\nabla_\beta(a_4 \phi_\lambda \phi^\lambda ) R^{\beta}{}_{\mu)\rho\sigma} \nb\\
 &&  -8 \varepsilon^{\alpha\beta\rho}{}_{(\nu} \nabla_\alpha(a_4 \phi_\lambda \phi^\lambda ) \nabla_\rho R_{\beta\mu)}.
\eqn

The expression of $B_{\mu\nu}$ in Eq.(\ref{field}) contains seven terms,
\bqn
B_{\mu\nu}=\sum_{A=1}^7 B_{\mu\nu}^{(A)},
\eqn
where
\bqn
B_{\mu\nu}^{(1)} &=&  \varepsilon^{\tau\lambda\alpha\beta} b_{1,X} \phi_{(\nu} \phi_{\mu)} R_{\alpha\beta\rho\sigma} \phi^\rho \phi_{\tau} \phi^\sigma_\lambda \nb\\
&& + \varepsilon^{\sigma\kappa\alpha\beta} b_1 R^{\rho}{}_{(\nu\alpha\beta} \phi_\rho \phi_\sigma \phi_{\mu)\kappa}  \nb\\
&&  -  \varepsilon^{\rho\beta\alpha}{}_{(\nu} \nabla_\sigma\nabla_\alpha \left( b_1 \phi_{\mu)} \phi_\rho \phi^\sigma_{\beta} \right) \nb\\
&& + \varepsilon^{\sigma\beta\alpha}{}_{(\nu} \nabla_\rho\nabla_\alpha \left(  b_1 \phi^\rho \phi_\sigma \phi_{\mu)\beta}  \right) \nb\\
&& -  \varepsilon^{\sigma\rho\alpha\beta} \nabla_\beta\nabla_\alpha \left( b_1 \phi_{(\nu} \phi_\sigma \phi_{\mu)\rho} \right) \nb\\
&&  - \frac{1}{2} \varepsilon^{\lambda}{}_{(\nu}{}^{\alpha\beta} \nabla_\sigma ( b_1 R^{\rho\sigma}{}_{\alpha\beta} \phi_\lambda \phi_{\mu)} ) \phi_\rho \nb\\
&&-  \frac{1}{2} \varepsilon^{\sigma\lambda\alpha\beta} \nabla_\lambda( b_1 R^{\rho}{}_{(\mu\alpha\beta} \phi_\rho \phi_{\nu)} ) \phi_\sigma \nb\\
&& +  \frac{1}{2} \varepsilon^{\sigma}{}_{(\nu}{}^{\alpha\beta} \nabla_\lambda (b_1 R^{\rho}{}_{\mu)\alpha\beta} \phi_\rho \phi_\sigma \phi^\lambda ) ,\\
B_{\mu\nu}^{(2)} &=&  \varepsilon^{\kappa\lambda\alpha\beta} b_{2,X}\phi_{(\nu} \phi_{\mu)} R_{\alpha\beta\rho\sigma} \phi^\rho_\kappa \phi^\sigma_\lambda \nb\\
&& + \varepsilon^{\sigma\kappa\alpha\beta} b_2 R^{\rho}{}_{(\nu\alpha\beta}  \phi_{\rho\sigma} \phi_{\mu)\kappa}  \nb\\
&&   -2 \varepsilon^{\rho\beta\alpha}{}_{(\nu} \nabla_\sigma \nabla_\alpha \left( b_2 \phi_{\mu)\rho} \phi^\sigma_{\beta} \right) \nb\\
&&  -  \varepsilon^{\sigma\rho\alpha\beta} \nabla_\beta\nabla_\alpha \left( b_2 \phi_{(\nu\sigma} \phi_{\mu)\rho} \right)   \nb\\
&& +  \varepsilon^{\lambda\alpha\beta}{}_{(\mu} \nabla_\rho  \left( b_2  R^{\rho\sigma}{}_{\alpha\beta} \phi_{\sigma\lambda}  \phi_{\nu)} \right) \nb\\
&& +  \varepsilon^{\rho\lambda\alpha\beta} \nabla_\rho \left( b_2  R^{\sigma}{}_{(\mu\alpha\beta} \phi_{\sigma\lambda}  \phi_{\nu)} \right) \nb\\
&& +  \varepsilon^{\rho\alpha\beta}{}_{(\mu} \nabla_\lambda \left(  b_2  R^{\sigma}{}_{\nu)\alpha\beta} \phi_{\sigma\rho}  \phi^\lambda \right)  ,\\
 B_{\mu\nu}^{(3)}&=&  \varepsilon^{\kappa\gamma\alpha\beta} b_{3,X} \phi_{(\mu} \phi_{\nu)} R_{\alpha\beta\rho\sigma} \phi^\sigma \phi^\rho_{\kappa} \phi^\lambda_{\gamma} \phi_\lambda \nb\\
&& + \varepsilon^{\sigma\kappa\alpha\beta} b_3 R^{\rho}{}_{(\nu\alpha\beta} \phi_{\mu)} \phi_{\rho\sigma} \phi^\lambda_{\kappa} \phi_\lambda \nb\\
&& + \varepsilon^{\gamma\lambda\alpha\beta} b_3 R^{\rho\sigma}{}_{\alpha\beta} \phi_\sigma \phi_{\rho \gamma} \phi_{(\mu \lambda} \phi_{\nu)} \nb\\
&&  - \varepsilon^{\rho\beta\alpha}{}_{(\nu} \nabla_\sigma \nabla_\alpha \left( b_3  \phi^\sigma \phi_{\mu)\rho} \phi^\lambda_{ \beta} \phi_\lambda \right) \nb\\
&&  - \varepsilon^{\sigma\rho\alpha\beta}\nabla_\beta \nabla_\alpha \left( b_3  \phi_{(\mu} \phi_{\nu)\sigma} \phi^\lambda_{\rho} \phi_\lambda \right) \nb\\
&& + \varepsilon^{\sigma\beta\alpha}{}_{(\nu}\nabla_\rho \nabla_\alpha \left( b_3  \phi_{\mu)} \phi^{\rho}_{\sigma} \phi^\lambda_{\beta} \phi_\lambda \right)  \nb\\
&&  - \frac{1}{2}  \varepsilon_{(\mu}{}^{\kappa\alpha\beta} \nabla_\rho \left( b_3 R^{\rho\sigma}{}_{ \alpha\beta} \phi_\sigma  \phi_{\lambda \kappa} \phi^\lambda  \phi_{\nu)}  \right) \nb\\
&& - \frac{1}{2} \varepsilon^{\rho\kappa\alpha\beta} \nabla_\rho \left( b_3 R_{(\mu}{}^{\sigma}{}_{ \alpha\beta} \phi_\sigma  \phi_{\lambda \kappa} \phi^\lambda  \phi_{\nu)}  \right)       \nb\\
&&  - \frac{1}{2}  \varepsilon^{\kappa}{}_{(\nu}{}^{\alpha\beta} \nabla_\lambda \left( b_3 R^{\rho\sigma}{}_{ \alpha\beta} \phi_\sigma \phi_{\rho\kappa} \phi^\lambda \phi_{\mu)}  \right) \nb\\
 && - \frac{1}{2}  \varepsilon^{\lambda\kappa\alpha\beta} \nabla_\kappa \left( b_3 R^{\rho\sigma}{}_{ \alpha\beta} \phi_\sigma \phi_{\rho\lambda} \phi_{(\mu} \phi_{\nu)} \right)      \nb\\
&&+ \varepsilon_{(\mu}{}^{\rho\alpha\beta} \nabla_\kappa \left( b_3 R_{\nu)}{}^{\sigma}{}_{ \alpha\beta} \phi_\sigma  \phi_{\lambda \rho} \phi^\lambda  \phi^\kappa  \right) \nb\\
&& + \frac{1}{2}  \varepsilon^{\lambda}{}_{(\nu}{}^{\alpha\beta} \nabla_\kappa \left( b_3 R^{\rho\sigma}{}_{ \alpha\beta} \phi_\sigma \phi_{\rho\lambda} \phi_{\mu)} \phi^\kappa  \right), \\
 B_{\mu\nu}^{(4)}&=& \varepsilon^{\kappa\gamma\alpha\beta} b_{4,X} \phi_{(\mu}\phi_{\nu)} R_{\alpha\beta\rho\sigma} \phi_{\gamma} \phi^\rho_{\kappa} \phi^\sigma_\lambda \phi^\lambda \nb\\
&&  + \varepsilon^{\sigma\kappa\alpha\beta} b_4 R^{\rho}{}_{(\nu\alpha\beta} \phi_\kappa \phi_{\rho\sigma} \phi_{\mu)\lambda} \phi^\lambda \nb\\
&&  + \varepsilon^{\gamma\lambda\alpha\beta} b_4 R^{\rho\sigma}{}_{\alpha\beta} \phi_\lambda \phi_{\rho \gamma} \phi_{\sigma(\nu} \phi_{\mu)}      \nb\\
&&  - \varepsilon^{\rho\beta\alpha}{}_{(\nu} \nabla_\sigma \nabla_\alpha \left( b_4 \phi_\beta \phi_{\mu)\rho} \phi^{\sigma}_{\lambda} \phi^\lambda \right) \nb\\
&& - \varepsilon^{\sigma\rho\alpha\beta} \nabla_\beta\nabla_\alpha \left( b_4\phi_\rho \phi_{(\nu\sigma} \phi_{\mu)\lambda} \phi^\lambda \right) \nb\\
&&  + \varepsilon^{\sigma\beta\alpha}{}_{(\nu} \nabla_\rho\nabla_\alpha \left( b_4 \phi_\beta \phi^{\rho}_{\sigma} \phi_{\mu)\lambda} \phi^\lambda \right) \nb \\
&&  - \frac{1}{2}   \varepsilon_{(\mu}{}^{\kappa\alpha\beta} \nabla_\rho \left( b_4 R^{\rho\sigma}{}_{\alpha\beta} \phi_\kappa \phi_{\sigma\lambda} \phi^\lambda \phi_{\nu)} \right) \nb\\
&& - \frac{1}{2}  \varepsilon^{\rho\kappa\alpha\beta} \nabla_\rho \left( b_4 R_{(\mu}{}^{\sigma}{}_{\alpha\beta} \phi_\kappa \phi_{\sigma\lambda} \phi^\lambda \phi_{\nu)} \right)  \nb\\
&&  - \frac{1}{2}  \varepsilon^{\kappa\lambda\alpha\beta} \nabla_\sigma \left( b_4 R^{\rho\sigma}{}_{\alpha\beta} \phi_\lambda \phi_{\rho\kappa} \phi_{(\nu} \phi_{\mu)} \right) \nb\\
&& - \frac{1}{2}  \varepsilon^{\sigma\kappa\alpha\beta} \nabla_\lambda \left( b_4 R^{\rho}{}_{(\mu\alpha\beta} \phi_\kappa \phi_{\rho\sigma} \phi^\lambda \phi_{\nu)} \right)   \nb\\
&&     + \frac{1}{2}  \varepsilon_{(\mu}{}^{\rho\alpha\beta} \nabla_\kappa \left( b_4 R_{\nu)}{}^{\sigma}{}_{\alpha\beta} \phi_\rho \phi_{\sigma\lambda} \phi^\lambda \phi^\kappa \right) \nb\\
&& + \frac{1}{2}  \varepsilon^{\sigma\lambda\alpha\beta} \nabla_\kappa \left( b_4 R^{\rho}{}_{(\mu\alpha\beta} \phi_\lambda \phi_{\rho\sigma} \phi_{\nu)} \phi^\kappa \right),\\
 B_{\mu\nu}^{(5)}&=& \varepsilon^{\kappa\gamma\alpha\beta} b_{5,X} \phi_{(\mu}\phi_{\nu)} R_{\alpha\rho\sigma\lambda} \phi^\rho \phi_\beta \phi^\sigma_{\kappa} \phi^\lambda_{\gamma} \nb\\
 && - 2\varepsilon^{\tau\sigma\alpha\beta} b_5 R^{\rho}{}_{\alpha(\nu\lambda} \phi_\rho \phi_\beta \phi_{\mu) \tau } \phi^\lambda_{\sigma}   \nb\\
&&  +\varepsilon^{\rho\lambda\alpha\beta}\nabla_\alpha\nabla_\sigma \left( b_5 \phi_{(\mu} \phi_\beta \phi^\sigma_{ \rho} \phi_{\nu)\lambda} \right) \nb\\
&&  + \varepsilon^{\alpha\rho}{}_{(\mu}{}^{\beta}\nabla_\lambda\nabla_\sigma \left( b_5 \phi_{\nu)} \phi_\beta \phi^\sigma_{ \alpha} \phi^\lambda_{ \rho} \right) \nb\\
&&  - \varepsilon^{\alpha\lambda}{}_{(\mu}{}^{\beta}\nabla_\rho\nabla_\sigma \left( b_5 \phi^\rho \phi_\beta \phi^\sigma_{\alpha} \phi_{ \nu)\lambda} \right) \nb\\
&& +  \varepsilon_{(\mu}{}^{\kappa\alpha\beta} \nabla_\sigma \left( b_5 R^{\rho}{}_{\alpha}{}^{\sigma\lambda} \phi_\rho \phi_\beta \phi_{\lambda \kappa} \phi_{\nu)}  \right)  \nb\\
&&+ \varepsilon^{\sigma\kappa\alpha\beta} \nabla_\sigma \left( b_5 R^{\rho}{}_{\alpha(\mu}{}^{\lambda} \phi_\rho \phi_\beta \phi_{\lambda \kappa} \phi_{\nu)}  \right) \nb\\
&& - \varepsilon_{(\mu}{}^{\sigma\alpha\beta} \nabla_\kappa \left( b_5 R^{\rho}{}_{\alpha\nu)}{}^{\lambda} \phi_\rho \phi_\beta \phi_{\lambda \sigma} \phi^\kappa\right) , \\
 B_{\mu\nu}^{(6)}&=& \ \varepsilon^{\kappa\tau\alpha\beta} b_{6,X} \phi_{(\mu}\phi_{\nu)}  R_{\beta\gamma} \phi_\alpha \phi^{\gamma}_{\kappa} \phi^{\lambda} _{\tau} \phi_\lambda \nb\\
 && +  \varepsilon^{\gamma \sigma\alpha\beta} b_6 R_{\beta(\mu} \phi_\alpha \phi_{\gamma\nu)} \phi^{\lambda}_{\sigma} \phi_\lambda \nb\\
&&  + \varepsilon^{\lambda\rho\alpha\beta} b_6 R_{\beta\gamma} \phi_\alpha \phi^{\gamma}_{\lambda} \phi_{\rho(\nu} \phi_{\mu)}  \nb\\
 &&   - \frac{1}{2} \varepsilon^{\tau\gamma\alpha\beta} \nabla_\beta \nabla_{(\mu} \left( b_6 g^{\lambda\rho} \phi_\alpha \phi_{\nu)\tau} \phi_{\rho\gamma} \phi_\lambda  \right) \nb\\
 && - \frac{1}{2} \varepsilon^{\beta\tau\alpha}{}_{(\mu}  \nabla_\gamma \nabla_{\nu)} \left( b_6 g^{\gamma \sigma} g^{\lambda\rho} \phi_\alpha \phi_{\sigma\beta} \phi_{\rho\tau} \phi_\lambda \right)   \nb\\
 &&     + \frac{1}{2} \varepsilon^{\beta\gamma\alpha}{}_{(\mu} \nabla_\tau \nabla_\kappa \left( b_6  g^{\lambda\rho} \phi_\alpha \phi_{\nu)\beta} \phi_{\rho\gamma} \phi_\lambda g^{\kappa\tau}  \right) \nb\\
&& + \frac{1}{2} \varepsilon^{\tau\kappa\alpha\beta} \nabla_\beta \nabla_\gamma \left(  b_6 g^{\gamma \sigma} g^{\lambda\rho} \phi_\alpha \phi_{\sigma\tau} \phi_{\rho\kappa} \phi_\lambda  g_{(\nu\mu)} \right) \nb\\
&&   + \frac{1}{2} \varepsilon^{\beta\tau\alpha}{}_{(\mu}  \nabla_{\nu)} \nabla_\gamma \left( b_6 g^{\gamma \sigma} g^{\lambda\rho} \phi_\alpha \phi_{\sigma\beta} \phi_{\rho\tau} \phi_\lambda  \right) \nb\\
&&  - \frac{1}{2}  \varepsilon^{\beta\kappa\alpha}{}_{(\mu} \nabla_{\nu)} \nabla_\gamma \left( b_6 g^{\gamma \sigma} g^{\lambda\rho} \phi_\alpha \phi_{\sigma\beta} \phi_{\rho\kappa} \phi_\lambda \right)  \nb\\
&&  -\frac{1}{2} \varepsilon^{\xi\sigma\alpha\beta}  \nabla_\xi \left( b_6 R_{\beta(\nu} \phi_\alpha g^{\lambda\rho} \phi_{\sigma\rho} \phi_\lambda \phi_{\mu)} \right) \nb\\
&&  -\frac{1}{2} \varepsilon_{(\mu}{}^{\xi\alpha\beta}  \nabla_\sigma \left( b_6 R_{\beta\gamma} \phi_\alpha g^{\gamma \sigma} g^{\lambda\rho} \phi_{\xi\rho} \phi_\lambda \phi_{\nu)} \right)   \nb\\
&&  +\frac{1}{2} \varepsilon_{(\mu}{}^{\sigma\alpha\beta}  \nabla_\xi \left( b_6 R_{\beta\nu)} \phi_\alpha g^{\lambda\rho} \phi_{\sigma\rho} \phi_\lambda \phi_\tau g^{\tau\xi} \right) \nb\\
&& -\frac{1}{2} \varepsilon^{\xi\rho\alpha\beta}\nabla_\rho \left(  b_6 R_{\beta\gamma} \phi_\alpha g^{\gamma \sigma} \phi_{\xi\sigma} \phi_{(\nu}\phi_{\mu)}  \right)  \nb\\
&&   -\frac{1}{2} \varepsilon^{\xi}{}_{(\nu}{}^{\alpha\beta}  \nabla_\rho  \left( b_6 R_{\beta\gamma} \phi_\alpha g^{\gamma \sigma} g^{\lambda\rho}\phi_{\xi\sigma} \phi_\lambda  \phi_{\mu)}  \right) \nb\\
&& +\frac{1}{2} \varepsilon^{\rho}{}_{(\nu}{}^{\alpha\beta}  \nabla_\xi \left( b_6 R_{\beta\gamma} \phi_\alpha g^{\gamma \sigma} \phi_{\rho\sigma} \phi_{\mu)}  \phi_\tau g^{\tau\xi}  \right) ,\\
 B_{\mu\nu}^{(7)}&=&   \varepsilon^{\tau\gamma\alpha\beta} b_{7,X} \phi_{(\mu}\phi_{\nu)}  R^{\rho}{}_{\sigma\alpha\beta} \phi_\rho \phi_{\tau} \phi_{\kappa \gamma} g^{\sigma\kappa} g^{\lambda\delta} \phi_{\lambda\delta} \nb\\
&& - \varepsilon^{\xi\beta\alpha}{}_{(\nu}  \nabla_\sigma \nabla_\alpha \left(  b_7 \phi_{\mu)} \phi_\xi \phi_{\kappa\beta} g^{\sigma\kappa} g^{\lambda\delta} \phi_{\lambda\delta} \right)  \nb\\
&& -  \varepsilon^{\sigma\xi\alpha\beta}  \nabla_\beta \nabla_\alpha \left( b_7 \phi_{(\nu} \phi_\sigma \phi_{\mu)\xi}  g^{\lambda\delta} \phi_{\lambda\delta} \right) \nb\\
&& +  \varepsilon^{\sigma\beta\alpha}{}_{(\nu}  \nabla_\xi \nabla_\alpha \left( b_7 \phi_\rho \phi_\sigma \phi_{\mu)\beta} g^{\lambda\delta} \phi_{\lambda\delta} g^{\rho \xi} \right) \nb\\
 && + \varepsilon^{\sigma\kappa\alpha\beta} b_7 R^{\rho}{}_{(\mu\alpha\beta} \phi_\rho \phi_\sigma \phi_{\nu)\kappa} g^{\lambda\delta} \phi_{\lambda\delta} \nb\\
&& + \varepsilon^{\lambda\tau\alpha\beta} b_7 R^{\rho}{}_{\sigma\alpha\beta} \phi_\rho \phi_\lambda \phi_{\kappa\tau} g^{\sigma\kappa} \phi_{(\mu\nu)}   \nb\\
&&  -\frac{1}{2}  \varepsilon^{\xi}{}_{(\nu}{}^{\alpha\beta} \nabla_\kappa \left( b_7 R^{\rho}{}_{\sigma\alpha\beta} \phi_\rho \phi_\xi  g^{\sigma\kappa} g^{\lambda\delta} \phi_{\lambda\delta} \phi_{\mu)} \right) \nb\\
&&  -\frac{1}{2}  \varepsilon^{\kappa\xi\alpha\beta} \nabla_\xi \left(  b_7 R^{\rho}{}_{(\mu\alpha\beta} \phi_\rho \phi_\kappa g^{\lambda\delta} \phi_{\lambda\delta} \phi_{\nu)} \right)  \nb\\
&&  +\frac{1}{2}  \varepsilon^{\kappa}{}_{(\nu}{}^{\alpha\beta} \nabla_\xi \left( b_7 R^{\rho}{}_{\mu)\alpha\beta} \phi_\rho \phi_\kappa  g^{\lambda\delta} \phi_{\lambda\delta} \phi_\gamma g^{\gamma \xi}  \right) \nb\\
&& - \frac{1}{2} \varepsilon^{\xi\delta\alpha\beta} \nabla_{(\nu} \left( b_7 R^{\rho}{}_{\sigma\alpha\beta} \phi_\rho \phi_\xi \phi_{\kappa\delta} g^{\sigma\kappa} \phi_{\mu)} \right)  \nb\\
&&   - \frac{1}{2} \varepsilon^{\lambda\xi\alpha\beta} \nabla_{(\mu} \left(  b_7 R^{\rho}{}_{\sigma\alpha\beta} \phi_\rho \phi_\lambda \phi_{\kappa\xi} g^{\sigma\kappa} \phi_{\nu)} \right) \nb\\
&& + \frac{1}{2} \varepsilon^{\lambda\delta\alpha\beta} \nabla_\xi \left( b_7 R^{\rho}{}_{\sigma\alpha\beta} \phi_\rho \phi_\lambda \phi_{\kappa\delta} g^{\sigma\kappa} g_{(\mu\nu)} \phi_\gamma g^{\gamma \xi} \right) . \nb\\
 \eqn

The expression of $F_{\phi}$ in Eq.(\ref{scalar_Eq}) is given by
\bqn
F_{\phi} = \sum_{A=1}^4 F_{a_A} + \sum_{B=1}^7 F_{b_B}, \lb{Fphi}
\eqn
where $F_{a_A}$ and $F_{b_B}$ are
\bqn
F_{a_1}
&=&   a_{1,\phi}\varepsilon^{\mu\nu\alpha \beta} R_{\alpha \beta \rho}{}^{\sigma} R_{\mu \nu}{}^{\rho\lambda}\phi_\sigma \phi_\lambda \nb\\
&&  - 2\varepsilon^{\mu\nu\alpha \beta}\nabla_m\left( a_{1,X} R_{\alpha \beta \rho}{}^{\sigma} R_{\mu \nu}{}^{\rho\lambda}\phi_\sigma \phi_\lambda\phi^m \right)  \nb\\
&&    - 2 \varepsilon^{\mu\nu\alpha \beta}\nabla_\sigma\left(a_1 R_{\alpha \beta \rho}{}^{\sigma} R_{\mu \nu}{}^{\rho\lambda}\phi_\lambda \right), \\
F_{a_2}
&=&    a_{2,\phi}  \varepsilon^{\mu\nu\alpha \beta} R_{\alpha \beta \rho \sigma} R_{\mu}{}^{ \lambda \rho \sigma}\phi_\sigma \phi_\lambda \nb\\
&&  - 2 \varepsilon^{\mu\nu\alpha \beta} \nabla_m\left( a_{2,X} R_{\alpha \beta \rho \sigma} R_{\mu}{}^{ \lambda \rho \sigma}\phi_\nu \phi_\lambda\phi^m \right)  \nb\\
&&   -   \varepsilon^{\mu\nu\alpha \beta} \nabla_\nu\left(a_2 R_{\alpha \beta \rho \sigma} R_{\mu}{}^{ \lambda \rho \sigma}\phi_\lambda \right) \nb\\
&& - \varepsilon^{\mu\nu\alpha \beta}\nabla_\lambda\left(a_2 R_{\alpha \beta \rho \sigma} R_{\mu}{}^{ \lambda \rho \sigma}\phi_\nu \right) , \\
F_{a_3}
&=&  a_{3,\phi} \varepsilon^{\mu\nu\alpha \beta} R_{\alpha \beta}{}^{ \rho}{}_{ \sigma} R^{\sigma}{}_{\nu}\phi_\rho \phi_\mu \nb\\
&& - 2  \varepsilon^{\mu\nu\alpha \beta} \nabla_m\left( a_{3,X} R_{\alpha \beta}{}^{ \rho}{}_{ \sigma} R^{\sigma}{}_{\nu}\phi_\rho \phi_\mu\phi^m \right)  \nb\\
&&  -  \varepsilon^{\mu\nu\alpha \beta} \nabla_\rho\left(a_3 R_{\alpha \beta}{}^{ \rho}{}_{ \sigma} R^{\sigma}{}_{\nu}\phi_\mu \right) \nb\\
&&  -  \varepsilon^{\mu\nu\alpha \beta} \nabla_\mu\left(a_3 R_{\alpha \beta}{}^{ \rho}{}_{ \sigma} R^{\sigma}{}_{\nu}\phi_\rho \right)  , \\
F_{a_4}
&=&   a_{4,\phi} \varepsilon^{\mu\nu\alpha \beta} R_{\rho\sigma \alpha \beta} R^{\alpha\beta}{}_{\mu \nu}\phi^\lambda \phi_\lambda \nb\\
&& - 2   \varepsilon^{\mu\nu\alpha \beta} \nabla_m\left( a_{4,X} R_{\rho\sigma \alpha \beta} R^{\alpha\beta}{}_{\mu \nu}\phi^\lambda \phi_\lambda\phi^m \right)  \nb\\
&&  - 2  \varepsilon^{\mu\nu\alpha \beta} \nabla_\lambda\left(a_4 R_{\rho\sigma \alpha \beta} R^{\rho\sigma}{}_{\mu \nu}\phi^\lambda \right) , \\
F_{b_1}
&=&    b_{1,\phi} \varepsilon^{\mu\nu\alpha \beta} R_{\alpha \beta}{}^{\rho\sigma } \phi_\rho\phi_\mu\phi_{\sigma\nu } \nb\\
&& -\varepsilon^{\mu\nu\alpha \beta} \nabla_\lambda\left( 2b_{1,X} R_{\alpha \beta}{}^{\rho\sigma } \phi^\lambda \phi_\rho\phi_\mu\phi_{\sigma\nu }\right)  \nb\\
&&    -\varepsilon^{\mu\nu\alpha \beta} \nabla_\rho \left(b_1 R_{\alpha \beta}{}^{\rho\sigma } \phi_\mu\phi_{\sigma\nu }\right) \nb\\
&&  -\varepsilon^{\mu\nu\alpha \beta} \nabla_\mu \left(b_1 R_{\alpha \beta}{}^{\rho\sigma } \phi_\rho\phi_{\sigma\nu }\right) \nb\\
&&  +\varepsilon^{\mu\nu\alpha \beta} \nabla_\nu\nabla_\sigma \left(b_1 R_{\alpha \beta}{}^{\rho\sigma }\phi_\rho\phi_\mu \right) , \\
F_{b_2}
&=&    b_{2,\phi} \varepsilon^{\mu\nu\alpha \beta} R_{\alpha \beta}{}^{\rho\sigma } \phi_{\rho\mu}\phi_{\sigma\nu } \nb\\
&& - \varepsilon^{\mu\nu\alpha \beta} \nabla_m \left( 2b_{2,X} R_{\alpha \beta}{}^{\rho\sigma } \phi^m \phi_{\rho\mu}\phi_{\sigma\nu }\right) \nb\\
&&   + 2\varepsilon^{\mu\nu\alpha\beta} \nabla_\mu\nabla_\rho\left(b_2  R_{\alpha \beta}{}^{\rho\sigma }  \phi_{\sigma\nu } \right) , \\
F_{b_3}
&=&    b_{3,\phi} \varepsilon^{\mu\nu \alpha \beta} R_{\alpha \beta \rho\sigma}\phi_\sigma \phi_{\rho\mu} \phi_{\lambda\nu} \phi^\lambda \nb\\
&& - 2 \varepsilon^{\mu\nu \alpha \beta} \nabla_m \left( b_{3,\phi} R_{\alpha \beta \rho\sigma}\phi^m \phi_\sigma \phi_{\rho\mu} \phi_{\lambda\nu} \phi^\lambda \right) \nb\\
&&   -  \varepsilon^{\mu\nu \alpha \beta} \nabla_\sigma \left( b_3 R_{\alpha \beta \rho\sigma} \phi_{\rho\mu} \phi_{\lambda\nu} \phi^\lambda \right) \nb\\
&& -  \varepsilon^{\mu\nu \alpha \beta} \nabla^\lambda\left( b_3 R_{\alpha \beta \rho\sigma} \phi_\sigma \phi_{\rho\mu} \phi_{\lambda\nu} \right) \nb\\
&&   +  \varepsilon^{\mu\nu \alpha \beta} \nabla_\mu\nabla_\rho\left( b_3 R_{\alpha \beta \rho\sigma} \phi_\sigma\phi^\lambda\phi_{\lambda\nu} \right) \nb\\
&&  + \varepsilon^{\mu\nu \alpha \beta} \nabla_\nu\nabla_\lambda\left( b_3 R_{\alpha \beta \rho\sigma} \phi_\sigma \phi^\lambda \phi_{\rho\mu} \right) , \\
F_{b_4}
&=&     b_{4,\phi} \varepsilon^{\mu\nu \alpha \beta} R_{\alpha \beta \rho\sigma} \phi_\nu \phi_{\rho\mu } \phi_{\sigma\lambda} \phi^\lambda \nb\\
&& - 2 \varepsilon^{\mu\nu \alpha \beta} \nabla_m \left( b_{4,\phi} R_{\alpha \beta \rho\sigma}\phi^m \phi_\nu \phi_{\rho\mu } \phi_{\sigma\lambda} \phi^\lambda \right) \nb\\
&&    - \varepsilon^{\mu\nu \alpha \beta} \nabla_\nu \left( b_4 R_{\alpha \beta \rho\sigma} \phi_{\rho\mu} \phi_{\sigma\lambda} \phi^\lambda \right) \nb\\
&&  -  \varepsilon^{\mu\nu \alpha \beta} \nabla_\lambda\left( b_4 R_{\alpha \beta \rho\sigma} \phi_\nu \phi_{\rho\mu} \phi_{\sigma}^{\lambda}  \right) \nb\\
&&    +  \varepsilon^{\mu\nu \alpha \beta} \nabla_\mu\nabla_\rho\left( b_4 R_{\alpha \beta \rho\sigma} \phi_\nu\phi^\lambda\phi_{\lambda}^{\sigma}\right) \nb\\
&& +  \varepsilon^{\mu\nu \alpha \beta} \nabla_\lambda\nabla_\sigma\left( b_4 R_{\alpha \beta \rho\sigma} \phi_\nu \phi^\lambda \phi_{\rho\mu} \right) , \\
F_{b_5}
&=&   b_{5,\phi} \varepsilon^{\mu\nu \alpha \beta} R_{\alpha}{}^{ \rho\sigma \lambda }\phi_\rho \phi_\beta \phi_{\sigma\mu} \phi_{\lambda\nu} \nb\\
&& - 2 \varepsilon^{\mu\nu \alpha \beta} \nabla_m \left( b_{5,\phi} R_{\alpha}{}^{ \rho\sigma \lambda }\phi^m \phi_\rho \phi_\beta \phi_{\sigma\mu} \phi_{\lambda\nu} \right) \nb\\
&&   -  \varepsilon^{\mu\nu \alpha \beta} \nabla_\rho \left( b_5 R_{\alpha}{}^{ \rho\sigma \lambda } \phi_\beta \phi_{\sigma\mu} \phi_{\lambda\nu} \right) \nb\\
&&  -  \varepsilon^{\mu\nu \alpha \beta} \nabla_\beta\left( b_5 R_{\alpha}{}^{ \rho\sigma \lambda } \phi_\rho \phi_{\sigma\mu} \phi_{\lambda\nu} \right) \nb\\
&&   +  \varepsilon^{\mu\nu \alpha\beta} \nabla_\mu\nabla_\sigma\left( b_5  R_{\alpha}{}^{ \rho\sigma \lambda } \phi_\rho \phi_\beta \phi_{\lambda\nu} \right) \nb\\
&& +  \varepsilon^{\mu\nu \alpha\beta} \nabla_\nu\nabla_\lambda\left( b_5 R_{\alpha}{}^{ \rho\sigma \lambda } \phi_\rho \phi_\beta \phi_{\sigma\mu} \right) , \\
F_{b_6}
&=&    b_{6,\phi} \varepsilon^{\mu\nu \alpha \beta} R_{\beta}{}^{ \gamma} \phi_\alpha \phi_{\gamma\mu} \phi_{\lambda\nu} \phi^\lambda \nb\\
&& - 2 \varepsilon^{\mu\nu \alpha \beta} \nabla_m \left( b_{6,\phi} R_{\beta}{}^{ \gamma} \phi^m \phi_\alpha \phi_{\gamma\mu} \phi_{\lambda\nu} \phi^\lambda \right) \nb\\
&&   -  \varepsilon^{\mu\nu \alpha \beta} \nabla_\alpha \left( b_6 R_{\beta}{}^{ \gamma} \phi_{\gamma\mu} \phi_{\lambda\nu} \phi^\lambda \right) \nb\\
&& -  \varepsilon^{\mu\nu \alpha \beta} \nabla_\lambda\left( b_6 R_{\beta}{}^{ \gamma} \phi_\alpha \phi_{\gamma\mu} \phi^{\lambda}_{\nu} \right) \nb\\
&&   + \varepsilon^{\mu\nu \alpha \beta} \nabla_\mu\nabla_\gamma\left( b_6 R_{\beta}{}^{ \gamma} \phi_\alpha \phi_{\lambda\nu} \phi^\lambda \right) \nb\\
&& +  \varepsilon^{\mu\nu \alpha \beta} \nabla_\nu\nabla_\lambda\left( b_6 R_{\beta}{}^{ \gamma} \phi_\alpha \phi_{\gamma\mu} \phi^\lambda \right) , \\
F_{b_7}
&=&    b_{7,\phi}  \varepsilon^{\mu\nu\alpha \beta} R_{\alpha \beta}{}^{\rho\sigma } \phi_\rho\phi_\mu\phi_{\sigma\nu } \nabla^2\phi \nb\\
&& - 2 \varepsilon^{\mu\nu\alpha \beta} \nabla_\lambda\left( b_{7,X} R_{\alpha \beta}{}^{\rho\sigma } \phi^\lambda \phi_\rho\phi_\mu\phi_{\sigma\nu }\nabla^2\phi\right)  \nb\\
&&  -  \varepsilon^{\mu\nu\alpha \beta} \nabla_\rho \left( b_7 R_{\alpha \beta}{}^{\rho\sigma } \phi_\mu\phi_{\sigma\nu }\nabla^2\phi \right) \nb\\
&&  -  \varepsilon^{\mu\nu\alpha \beta} \nabla_\mu \left( b_7 R_{\alpha \beta}{}^{\rho\sigma } \phi_\rho\phi_{\sigma\nu }\nabla^2\phi \right)  \nb\\
&&  +  \varepsilon^{\mu\nu\alpha \beta} \nabla_\nu\nabla_\sigma \left( b_7 R_{\alpha \beta}{}^{\rho\sigma }\phi_\rho\phi_\mu \nabla^2\phi \right) \nb\\
&& + \varepsilon^{\mu\nu\alpha \beta} \nabla^2 \left( b_7 R_{\alpha \beta}{}^{\rho\sigma } \phi_\rho\phi_\mu \phi_{\sigma\nu } \right).
\eqn


\end{document}